\documentclass[reqno]{article}

\usepackage{amsmath}
\usepackage{amsfonts}
\usepackage{graphics}
\usepackage{epsfig}
\usepackage{amssymb}
\usepackage{amsthm}
\usepackage{amscd}
\usepackage[all]{xy}
\usepackage{latexsym}
\usepackage{graphicx}
\usepackage{multirow}
\usepackage{geometry}
\usepackage{color}
\usepackage{CJKutf8} 
\usepackage{bm}
\usepackage{mathrsfs}

    \theoremstyle{definition}
    \newtheorem{dfn}{Definition}[section]
    \newtheorem{prop}[dfn]{Proposition}
    
    \newtheorem{thm}[dfn]{Theorem}
    
    \newtheorem{conj}[dfn]{Conjecture}
    \newtheorem{rem}[dfn]{Remark}
    \newtheorem{ass}[dfn]{Assumption}
    \newtheorem{ex}[dfn]{Example}
    \newtheorem{exc}{Exercise}
    \newtheorem*{ack}{Acknowledgement}

\numberwithin{equation}{section}
\numberwithin{figure}{section}
\numberwithin{table}{section}

\def\rchi{{\hbox{\raise1.5pt\hbox{$\chi$}}}}

\newcommand{\bea}{\begin{eqnarray}}
\newcommand{\eea}{\end{eqnarray}}
\newcommand{\be}{\begin{equation}}
\newcommand{\ee}{\end{equation}}

\newcommand{\Res}{\mathop{\rm Res}}

\begin{document}
\large
\setcounter{section}{0}

\title{\huge  Les Houches Lectures on \\ Exact WKB Analysis and Painlev\'e Equations}

\large

\author{Kohei Iwaki \quad (The Univeristy of Tokyo)}

\date{}







\maketitle

\begin{abstract}
The first part of these lecture notes is devoted to an introduction to the theory of 
exact WKB analysis for second-order Schr\"odinger-type ordinary differential equations. 
It reviews the construction of the WKB solution, Borel summability, connection formulas, 
and their application to direct monodromy problems. 
In the second part, we discuss recent developments in applying exact WKB analysis 
to the study of Painlev\'e equations. By combining exact WKB analysis with topological recursion, 
it becomes possible to explicitly compute the monodromy of linear differential equations associated 
with Painlev\'e equations, assuming Borel summability and other conditions. 
Furthermore, by using isomonodromy deformations (integrability of the Painlev\'e equations), 
the resurgent structure of the $\tau$-function and partition function is analyzed.

These lecture notes accompanied a series of lectures at the Les Houches school, 
``Quantum Geometry (Mathematical Methods for Gravity, Gauge Theories and Non-Perturbative Physics)'' 
in Summer 2024.

\end{abstract}

\allowdisplaybreaks

\tableofcontents

\setlength{\parskip}{0.1ex}

\setcounter{section}{-1}

\section{Introduction}

Both {\em exact WKB analysis} and {\em Painlev\'e equations} 
have been significant topics in both mathematics and physics.
The purpose of this lecture note is to provide an overview of 
the theory of exact WKB analysis for Schr\"odinger-type 
ordinary differential equations (ODEs),  
and to outline recent results, including several conjectures, 
regarding its applications to the Painlev\'e equations.

Exact WKB analysis 
is a powerful method for studying Schr\"odinger-type ODEs 
(i.e., second-order linear ODEs with a small parameter $\hbar$) 
defined on a complex domain. 
From a mathematical perspective, 
it serves as a tool to analyze singularly perturbed differential equations.
The classical WKB (Wentzel--Kramers--Brillouin) method typically provides 
an approximate solution to linear differential equations in the semi-classical regime, 
where the parameter $\hbar$ is considered small. 
However, it is well-known that even when higher-order correction terms in 
$\hbar$ are included in the WKB approximation to construct an exact solution, 
the resulting series often diverges. 
Nevertheless, this formal solution can be promoted to an exact solution using the Borel summation method.
Through the analysis of the singularities of the Borel transform, 
the Stokes phenomenon with respect to $\hbar$ is analyzed, 
leading to an explicit {\em connection formula} that describes 
the analytic continuation of the Borel-resummed WKB solution with respect to the independent variable. 

The foundational idea of exact WKB analysis was presented in the celebrated paper \cite{Voros83} by Voros, 
and it was applied to the eigenvalue problems in quantum mechanics (\cite{DDP93, DDP97-a, DDP97-b}). 
It was later brought to Japan by Sato, 
and has been studied from the perspective of {\em algebraic analysis} 
by Aoki, Kawai, Takei (\cite{SAKT91, AKT91, KT98}). 
A notable outcome of their work is the application of exact WKB analysis 
to the calculation of {\em monodromy matrices} for Fuchsian Schr\"odinger-type ODEs.
In these works, it has been shown that the monodromy matrices of the exact solution, 
obtained by Borel summing the WKB solution, are expressed in terms of 
a generating series of period integrals on the spectral curve (\cite[\S 3]{KT98}).
The generating series of these periods is referred to as the {\em Voros period}\footnote{
There are several names for this object. 
It is also called {\em Voros coefficients (multipliers)}, 
{\em quantum period}, {\em all-order Bohr--Sommerfeld period}, 
(logarithm of) {\em Fock--Goncharov coordinates}, (logarithm of) {\em spectral coordinates}, etc. }. 
Although this note cannot cover them, the Voros period plays an important role 
not only in exact WKB analysis but also in 
its applications to related topics in mathematical physics;
see \cite{All18, AT13, Bri16, CLT20, HK17, HN13, HK17, HN19, 
IK-I, IK-II, IKoT-1, IKoT-2, IN14, IN15, KMSU23, Sue, Takei08} for example.

Independently of the above developments, in \cite{GMN09, GMN12}, 
Gaiotto, Moore, and Neitzke discovered a beautiful relationship between 
the {\em BPS indices} of $4d$ ${\mathcal N} = 2$ supersymmetric field theory and exact WKB analysis. 
In their work, the Stokes graph in exact WKB analysis is understood as the {\em spectral network}, 
through which the wall-crossing formulas for BPS indices are analyzed. 
This discovery revealed new aspects of exact WKB analysis and has led to numerous subsequent applications. 
In fact, some of the applications mentioned above were inspired by these results.
While we confine ourselves to brief remarks in \S \ref{subsec:comments-wkb} of this note, 
we refer the reader to the review \cite{KKRTT24} for further details.

On the other hand, Painlev\'e equations are a class of second-order nonlinear ODEs that 
were discovered in the early 20th century by Painlev\'e (\cite{Painleve}). 
These equations are distinguished by a property known as the {\em Painlev\'e property}, 
which ensures that their solutions have no movable branch points or movable essential singularities (\cite{Conte}). 
The six Painlev\'e equations, labeled Painlev\'e I through Painlev\'e VI, 
exhibit remarkable features such as affine-Weyl symmetry, 
the geometry of the space of initial conditions, and more
(\cite{Boalch07, Boalch20, Boa-Yam15, Hiroe17, IIS03, IIS06, Iwasaki91,
Okamoto79, Okamoto-studies, Okamoto-99-ham, OKSO06, Sakai00}). 
Painlev\'e equations frequently appear in various areas of mathematical physics; for instance, see
\cite{ASV11, BGMT24, BLMST, BK90, CLT20, DS90, dubrovin99, GIL, GIKM, ILT, JMMS80, SMJ-78-80, WMTB76, TW94}.
Currently, various extensions such as higher-order Painlev\'e equations, 
($q$-)difference Painlev\'e equations, and elliptic Painlev\'e equations are known; 
see \cite{HKNS18, Joshi-book, NY98, Sakai00} for example.
Due to their rich structures and connections to diverse fields, 
the Painlev\'e equations remain an active area of research in both mathematics and physics.

Among the properties of Painlev\'e equations, 
the description through {\em isomonodromy deformations} (\cite{JMU, Okamoto-99-im}) 
is particularly important in this note.
This means that the solutions of the Painlev\'e equations possess 
the monodromy matrix (or Stokes matrix) of a certain linear ODE as conserved quantities. 
The existence of such conserved quantities reflects the {\em integrability} of the Painlev\'e equations, 
aiding in the study of nontrivial properties of their solutions, such as the {\em nonlinear connection problem}. 
For previous works in this direction, see \cite{Kap89, FIKN, Kitaev, LR} for example.
As mentioned earlier, since exact WKB analysis has proven effective in analyzing the monodromy of linear ODEs, 
it is natural to explore its application to the Painlev\'e equations through isomonodromy deformations. 
In Part II, we introduce this approach using the Painlev\'e I equation as an example 
(but we expect that the same method can be applied to other Painlev\'e equations as well).
We also note that, from a similar perspective, Aoki, Kawai, and Takei 
pioneered the theory of exact WKB analysis for the Painlev\'e equations
(\cite{KT96, AKT96, KT98-Painleve, Takei98}; see also \cite[\S 4]{KT98}).
Although our approach differs from previous studies, 
the results of \cite{I19}, which are fundamental in Part II, 
are strongly influenced by the Aoki--Kawai--Takei theory.

In our approach, the {\em topological recursion} (\cite{CEO06, EO07}) 
also plays a crucial role. 
Topological recursion is linked to exact WKB analysis via the theory of 
{\em quantum curve} (\cite{BouE, DM14, GS}).
Moreover, it is known that the $\tau$-function of the Painlev\'e equations can be constructed as 
(the discrete Fourier transform of) the {\em partition function} of the topological recursion; 
see \cite{IM, IMS, IS} for perturbative case, and 
\cite{EG, EGFMO, I19, MO} for non-perturbative case. 
Although we will not discuss it in detail, the {\em Kyiv formula} (\cite{GIL}), 
which constructs the Painlev\'e $\tau$-function 
using the conformal blocks, is closely related to the non-perturbative case.
Through the topological recursion, the relationship between 
exact WKB analysis and the Painlev\'e equations becomes even more robust.

However, as explained in Part II, the theorems reviewed in Part I cannot be directly applied 
to isomonodromic linear ODE associated with Painlev\'e equation. 
We will explicitly identify the issues and propose specific conjectures (Conjecture \ref{conj:hope})
that are expected to hold in order to resolve those problems.
Assuming these conjectures, the application of exact WKB analysis enables 
an explicit description of the Stokes multipliers of 
the isomonodromic linear ODE in terms of Voros periods (\cite[\S 5]{I19}). 
Furthermore, by utilizing the invariance of the Stokes multipliers, 
a characteristic property of the Painlev\'e equations, 
it follows that the {\em Delabere--Dillinger--Pham formula} (\cite{DDP93}) 
in exact WKB analysis describes the non-linear connection formula for the Painlev\'e $\tau$-function. 
This derivation is inspired by the considerations in \cite{Takei95, Takei98}, 
and see also \cite{BK19, CLT20, BSSV22} for further references.
These, in turn, yield an explicit conjectural formula concerning the {\em resurgent structure} 
of the partition functions for the topological recursion (\cite{IM23}). 
Interestingly, the conjectural formula derived based on the properties of 
the Painlev\'e equations coincides exactly with a conjectural formula 
derived from the non-perturbative analysis of the {\em holomorphic anomaly equation} 
in topological string theory obtained recently (\cite{CSESV14, CSESV13, gkkm23, gm22}). 

These considerations suggest that the integrability of the Painlev\'e equations, 
as derived through isomonodromy deformations, 
governs the {\em non-perturbative effects} (Stokes phenomena) 
in topological recursion and topological string theory. 
This adds a new perspective to the relationship between Painlev\'e equations and mathematical physics. 
The goal of Part II is to present these intriguing applications in cutting-edge topics in mathematical physics.
While initially proposed by Painlev\'e as candidates for ``new special functions'', 
over a century later, the solutions of Painlev\'e equations are gaining even greater importance 
as special functions. 

\bigskip
This lecture note is organized as follows. 
In Part I, we first recall known results on the exact WKB analysis for Schr\"odinger-type linear ODEs. 
Starting with the construction of WKB solutions, we discuss methods for 
determining Borel summability based on Stokes graphs and describe the Voros/Koike connection formulas. 
We illustrate through concrete examples how the connection matrix can be explicitly expressed in terms of the Voros period.
Additionally, for the convenience of readers, we have decided to touch upon the outline 
of the proof of Borel summability by Koike--Sch\"afke \cite{Koike-Schafke}. 
This note is based on the foundational literature on exact WKB analysis by Kawa--Takei \cite{KT98}. 
We will also discuss on the analysis of turning points of simple-pole type 
as developed by Koike \cite{Koike2000}.

In Part II, we focus on the Painlev\'e I equation and introduce our approach based 
on the exact WKB analysis. We begin with a review of the description of Painlev\'e I 
via isomonodromy deformation and, following \cite{I19}, 
explain the construction of the $\tau$-function as well as the method for 
computing Stokes multipliers of the isomonodromic linear ODE based on conjectures inspired by exact WKB analysis. 
Finally, we discuss the conjectural formula regarding the resurgent structure of the partition function obtained in 
\cite{IM23}. We also provide a minimal explanation of the topological recursion which is essential 
in the construction of the $\tau$-function. 

At the end of both Part I and Part II, we briefly introduce several topics 
closely related to the respective contents. 
Many challenging problems remain in this field, 
and we look forward to the contributions of researchers entering this area.

\newpage
\section{Part I : Introduction to Exact WKB Analysis}

In the first part, we consider the Sch\"odinger-type ODEs of the form 
\begin{equation} \label{eq:sch}
\left( \hbar^2 \frac{{\rm d}^2}{{\rm d}x^2} - Q(x) \right) \psi(x,\hbar) = 0,
\end{equation}
where $\hbar$ is a parameter and $Q(x)$ is a rational function of $x$.  
The goal of Part I is to provide an explanation of the fundamental concepts 
of the exact WKB analysis (\cite{Voros83, KT98}) and to introduce its applications. 
In particular, we will see how the monodromy or Stokes matrix of equation \eqref{eq:sch}
is described using the Voros periods. 
This point of view will be quite useful in studying the Painlev\'e equation in Part II.

We denote by  
\begin{equation} \label{eq:phi}
\phi(x) = Q(x) {\rm d}x^{2}
\end{equation}
the meromorphic quadratic differential on ${\mathbb P}^1$ associated with \eqref{eq:sch}. 
Below we assume:
\begin{ass} ~ \label{ass:part1}
\begin{itemize}
\item 
$\hbar$ is a real positive and small parameter; that is, 
$0 < \hbar \ll 1$.
\item 
$\phi$ has at least one zero or one simple pole. 
\item 
All zeros of $\phi$ are simple.
\item 
$\phi$ has a pole at $x=\infty$ of order at least $2$. 
\end{itemize}
\end{ass}
We note that the quadratic differentials satisfying the GMN condition in the sense of 
\cite[Definition 2.1]{BS} always satisfy the last three conditions in Assumption \ref{ass:part1}.

\subsection{WKB solution, spectral curve and turning points}

Here we recall a construction of (formal) WKB solutions of 
the Schr\"odinger-type ODE \eqref{eq:sch} along the discussion in \cite[\S 2]{KT98}. 
 
\subsubsection{Construction of WKB solution}
\label{subsec:wkb-recursion}

Let us take a new unknown function $P(x,\hbar)$ defined by 
\begin{equation}
\psi(x,\hbar) = \exp\left( \int^{x}_{x_0} P(x', \hbar) \, {\rm d}x' \right)
\end{equation}
with $x_0$ being a generic base point. 
Then, it satisfies the so-called Riccati equation 
\begin{equation} \label{eq:riccati}
\hbar^2 \left( P(x,\hbar)^2 + \frac{{\rm d}P}{{\rm d}x}(x,\hbar) \right) = Q(x)
\end{equation}
associated with \eqref{eq:sch}. 
Now we put the following WKB-ansatz:  
\begin{equation}
P(x,\hbar) = \sum_{m \ge -1} \hbar^{m} P_m(x) 
\end{equation}
where $P_m$ are assumed to be holomorphic on some domain. 
The Riccati equation \eqref{eq:riccati} then leads to 
the following set of recursion relations 
for $P_m$, which we call the {\em WKB recursion}: 
\begin{equation} \label{eq:P-1}
P_{-1}(x)^2 = Q(x), 
\end{equation}
\begin{equation} \label{eq:P0}
2 P_{-1}(x) P_0(x) + \frac{{\rm d}P_{-1}(x)}{{\rm d}x} = 0, 
\end{equation}
\begin{equation} \label{eq:Pm}
2 P_{-1}(x) P_{m+1}(x) + \sum_{\ell = 0}^{m} P_{\ell}(x) P_{m - \ell}(x) 
+ \frac{{\rm d}P_{m}(x)}{{\rm d}x} = 0 \qquad ( m \ge 0).
\end{equation}
We can easily solve the recursion relation term-by-term and obtain two solutions 
\begin{equation} \label{eq:coeff-pm}
P^{(\pm)}_{-1}(x) = \pm \sqrt{Q(x)}, \quad
P^{(\pm)}_{0}(x) = -\frac{Q'(x)}{4Q(x)}, \quad 
P^{(\pm)}_{0}(x) = \pm \frac{4 Q(x) Q''(x) - 5 (Q'(x))^2}{32 Q(x)^{5/2}}, \quad \dots 
\end{equation}
depending on the choice of the square root of $Q$.
Thus we obtain two formal solutions of the Riccati equation \eqref{eq:riccati} 
and the corresponding WKB solutions as
\begin{equation} \label{eq:p-and-psi}
P^{(\pm)}(x,\hbar) = \sum_{m \ge -1} \hbar^{m} P^{(\pm)}_m(x),
\quad
\psi_\pm (x,\hbar) = 
\exp\left( \int^{x}_{x_0} P^{(\pm)}(x',\hbar) \, {\rm d}x' \right).
\end{equation}
Here and in what follows, integrals of formal series are understood as term-wise integrals. 

It is convenient to use odd/even decomposition 
\begin{equation} \label{eq:odd-even}
P_{\rm odd}(x,\hbar) = \frac{P^{(+)}(x,\hbar) - P^{(-)}(x,\hbar)}{2}, \quad
P_{\rm even}(x,\hbar) = \frac{P^{(+)}(x,\hbar) + P^{(-)}(x,\hbar)}{2}.
\end{equation}
Then, we have $P^{(\pm)}(x,\hbar) = \pm P_{\rm odd}(x,\hbar) + P_{\rm odd}(x,\hbar)$
and 
\begin{equation}
P_{\rm even}(x, \hbar) = - \frac{1}{2 P_{\rm odd}(x,\hbar)} \frac{{\rm d}P_{\rm odd}}{{\rm d}x}(x,\hbar)
\end{equation}
holds as formal series in $\hbar$. Therefore, we can take 
\begin{equation} \label{eq:wkb-sol}
\psi_\pm (x,\hbar) = \frac{1}{\sqrt{P_{\rm odd}(x,\hbar)}} 
\exp\left( \pm \int^{x}_{x_0} P_{\rm odd}(x', \hbar) \, {\rm d}x'  \right)
\end{equation} 
as the WKB solution, which agrees with \eqref{eq:p-and-psi} up to
formal series of $\hbar$ with $x$-independent coefficients.
It can also be written as a formal series with an exponential factor: 
\begin{equation} \label{eq:wkb-series}
\psi_\pm(x,\hbar) = 
\exp\bigl( \pm \hbar^{-1} S(x) \bigr)
\sum_{m \ge 0} \hbar^{m + \frac{1}{2}} \psi_{\pm, m}(x), 
\quad 
S(x) = \int^{x}_{x_0} \sqrt{Q(x')}\,{\rm d}x'.
\end{equation}
If we truncate the series 
at the first term, 
then we have the traditional WKB approximation 
$\exp(\pm \hbar^{-1} S(x)) \, {Q(x)^{-1/4}}$.
The WKB solution \eqref{eq:wkb-sol} is normalized by specifying the lower endpoint $x_0$ and the path of integration. 
We will later discuss how to choose the normalization when describing connection formulas.

\begin{ex}
\label{ex:airy}
The Schr\"odinger-type ODE with $Q(x) = x$, that is, 
\begin{equation} \label{eq:airy-eq}
\left( \hbar^2 \frac{{\rm d}^2}{{\rm d}x^2} - x \right) \psi(x,\hbar) = 0
\end{equation}
is called the {\em Airy equation}. 
For this case, we have
\begin{equation}
P^{(\pm)}_{-1}(x) = \sqrt{x}, \quad P^{(\pm)}_{0}(x) = - \frac{1}{4x}, \quad
P^{(\pm)}_{1}(x) = \mp \frac{5}{32 x^{5/2}}, \quad
P^{(\pm)}_{2}(x) = - \frac{15}{64 x^{4}}, \quad
 \dots
\end{equation}
and 
\begin{equation} \label{eq:wkb-airy}
\psi_\pm^{\rm Airy}(x,\hbar) = \exp\left( \pm \hbar^{-1} \, \frac{2x^{3/2}}{3} \right) \frac{ \hbar^{1/2} }{x^{{1}/{4}}}
\left(1 \pm \frac{5}{48 x^{3/2}} \hbar + \frac{385}{4608 x^{3}} \hbar^2 + \cdots \right), 
\end{equation}
where we have taken the lower endpoint $x_0$ in \eqref{eq:wkb-sol} 
at $\infty$ (except for the leading term). 
It is known that the rational numbers appearing in the coefficients
are related to the intersection numbers on the moduli space 
${\mathcal M}_{g,n}$ of Riemann surfaces through the so-called 
{\em topological recursion/quantum curve (TR/QC) correspondence} (see \cite{GS, Zhou12, DM14, BouE}). 
We will also make several remarks about the 
application of the TR/QC correspondence in \S \ref{subsec:comments-wkb}.

\end{ex}

\begin{rem} \label{rem:hbar-correction}
The above construction of WKB solutions can be easily generalized 
to the case where the Sch\"odinger potential $Q(x)$ in \eqref{eq:sch} is a formal series in $\hbar$ of the form
\[
Q(x) = \sum_{m \ge 0} \hbar^{m} Q_m(x),
\]
by replacing the right hand side of \eqref{eq:P-1}, \eqref{eq:P0}, \eqref{eq:Pm} by
$Q_0(x)$, $Q_1(x)$, $Q_{m+2}(x)$, respectively\footnote{ 
In this case, $P_{\rm odd}$/$P_{\rm even}$ defined in \eqref{eq:odd-even}
can contain even/odd degree terms of $\hbar$, respectively.}. 
In the case, the associated quadratic differential \eqref{eq:phi} 
is replaced by $Q_0(x) {\rm d}x^{2}$ (see Exercise \ref{exe:oper} below), 
and we need further conditions to guarantee the Borel summability of WKB solutions 
in addition to Assumption \ref{ass:part1} (c.f., \cite[Assumption 2.5]{IN14}).
\end{rem}

\subsubsection{Spectral curve and turning points}

In view of \eqref{eq:coeff-pm}, it is found that each coefficient of 
$P^{(\pm)}$ 
is (possibly) a multivalued function of $x$.  
Therefore, it is natural to consider the following.

\begin{dfn}
The Riemann surface
\begin{equation}
\Sigma = \{(x,y) \in {\mathbb C}^2 ~|~ y^2 = Q(x) \}
\end{equation}
is called the {\em (WKB) spectral curve}, 
or the {\em classical limit} of the Schr\"odinger-type ODE \eqref{eq:sch}.
We also denote by $\overline{\Sigma}$ its compactification. 
\end{dfn}

The Riemann surface $\Sigma$ is an important geometric object in WKB analysis. 
We denote by $\pi$ the projection map $\overline{\Sigma} \ni (x,y) \mapsto x \in {\mathbb P}^1$. 
We usually identify the point $x$ with a point $(x,y)$ on $\Sigma$ by taking an appropriate branch cut
for the square root $\sqrt{Q(x)}$, and regard $x$ as a local coordinate of $\Sigma$ as well.

Note that the path of integration in \eqref{eq:p-and-psi} or \eqref{eq:wkb-sol} 
should be considered on 
\begin{equation}
\Sigma' = \Sigma \setminus \pi^{-1} ({\rm Crit}(\phi)), 
\end{equation}
where 
${\rm Crit}(\phi) = \{ \text{zeros and poles of $\phi$} \}$
is the set of critical points of $\phi$. 
These points also play crucial role in the WKB analysis.

\begin{dfn}
A {\em turning point} of the Schr\"odinger-type ODE \eqref{eq:sch}
is either zero or simple pole\footnote{The traditional WKB analysis 
only considered the zeros of $\phi$ as turning points, 
but Koike's research (\cite{Koike2000}) revealed that simple poles 
also play a similar role as turning points. 
Therefore, the above definition is now adopted. 
To distinguish it from the conventional turning points, 
it is sometimes referred to as a ``turning point of simple pole-type''. 
See also \cite{HN16}.
} 
of $\phi$. 
The poles of $\phi$ are called {\em singular points} 
of \eqref{eq:sch}.
\end{dfn}

The WKB solutions cannot be defined not only at singular points but also at turning points. 
In quantum mechanics, the connection problem of WKB solutions around the turning point 
has been used to study the quantization conditions. 
Such connection formulas will be discussed in \S \ref{sec:conn-formula} 
from the viewpoint of the Stokes phenomenon with respect to $\hbar$.

Before ending this subsection, we note that the WKB solutions thus 
constructed are usually factorially divergent series of $\hbar$ at any point $x$. 

\begin{prop} \label{prop:gevrey-1}
Let $K \subset {\mathbb C}$ be an arbitrary compact set that includes neither turning points nor singular points. 
Then, there exists $C_K, r_K > 0$ such that 
\begin{equation} \label{eq:gevrey-1}
\sup_{x \in K} |P^{(\pm)}_m(x)| \le C_K \, r_K^m \, m!
\end{equation} 
holds for all $m \ge 0$. 
\end{prop}

See \cite{AKT91} for a proof of this fact. 
In next subsection, we will give a criterion to determine 
when this divergent series can be Borel summable.

\bigskip
\begin{exc} \label{exe:oper}
Prove that the Schr\"odinger-type ODE \eqref{eq:sch} 
is transformed to 
\begin{equation}\label{eq:Sch-z-coord}
\left( \hbar^2 \frac{{\rm d}^2}{{\rm d}z^2} - \widetilde{Q}(z) \right) \widetilde{\psi}(z,\hbar) = 0, 
\quad
\widetilde{Q}(z) = \left(\frac{{\rm d}x(z)}{{\rm d}z}\right)^2 Q(x(z)) 
- \frac{\hbar^2}{2} \{x(z); z\}
\end{equation}
under a holomorphic coordinate change $x=x(z)$ 
combined with a gauge transformation 
$\widetilde{\psi} = ({\rm d}x/{\rm d}z)^{-1/2} \psi$.
Here, 
\begin{equation}
\{x(z); z\} = \frac{x'''(z)}{x'(z)} - \frac{3}{2} \left(\frac{x''(z)}{x'(z)} \right)^2
\end{equation} 
is the Schwarzian derivative\footnote{ 
\label{footnote:projective-connection}
Such objects that include the Schwarzian derivative in the transformation law 
are called {projective connections} (or {$sl_2$-opers}). 
}. 
Observe that, in the transformation law \eqref{eq:Sch-z-coord}, 
the leading term $Q_0$ of the potential function precisely satisfies 
the transformation rule for quadratic differentials. 
Prove also that the formal series $P_{\rm odd}$ transforms as follows:
\begin{equation}
\widetilde{P}_{\rm odd}(z, \hbar) = \frac{{\rm d}x(z)}{{\rm d}z} P_{\rm odd}(x(z), \hbar).
\end{equation}
\end{exc}

\smallskip
\begin{exc} \label{exc:pole-order}
Let $p$ be a pole of $\phi$ of order $r$ ($ \ge 1$). 
Verify that $P_m$ behaves as 
\begin{equation}
P_m(x) \sim \begin{cases}
\displaystyle c_m \, (x-p)^{\frac{(r-2)m - 2}{2}} 
& \text{for $p \in {\mathbb C}$},  \\[+.5em]
\displaystyle c_m \,  x^{- \frac{(r-2)m + 2}{2}} 
& \text{for $p = \infty$}
\end{cases}
\end{equation}
when $x \to p$, with some constants $c_m \in {\mathbb C}$.




\end{exc}

\smallskip
\begin{exc}[{c.f., \cite[Theorem 1.1]{Takei08}}] \label{eq:Weber-Bernoulli}
The Schr\"odinger-type ODE 
\begin{equation} \label{eq:Weber-eq}
\left( \hbar^2 \frac{{\rm d}^2}{{\rm d}x^2} - \Bigl( \frac{x^2}{4} - \nu \Bigr) \right) \psi(x,\hbar) = 0
\qquad(\nu \in {\mathbb C}^\ast)
\end{equation}
is called the {\em Weber equation}. 
For $m \ge 1$, compute the integral of $P_m$ along a path connecting 
two pre-images $\infty_\pm \in \overline{\Sigma}$ of $x = \infty$ by $\pi$ 
(see Figure \ref{fig:weber-path}), and check that 
\begin{equation} \label{eq:weber-voros-terms}
\int^{\infty_+}_{\infty_-} P_m(x) \, {\rm d}x 
= 
\begin{cases}
\dfrac{(2^{1-2k}-1) B_{2k}}{2k(2k-1) \nu^{2k-1}} 
& \text{if $m = 2k-1$ is odd} \\[+1.em]
0 &  \text{if $m = 2k$ is even}, 
\end{cases} 
\end{equation}
hold (up to overall sign). 
Here, $B_{2k}$ is the Bernoulli number defined by 
\begin{equation} \label{eq:def-bernoulli}
\frac{t}{{\rm e}^{t} - 1} = 1 - \frac{t}{2} + \sum_{k \ge 1} \frac{B_{2k}}{(2k)!} t^{2k}. 
\end{equation}

\vspace{-.7em}

 \begin{figure}[h]
  \begin{center} 
  \includegraphics[width=35mm]{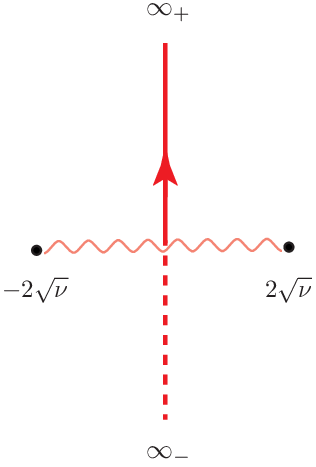} 
  \end{center} 
  \caption{A path from $\infty_-$ to $\infty_+$. 
  Here and in what follows, the wiggly lines represent branch cuts for $\sqrt{Q(x)}$. 
  The solid part (resp., the dotted part) of the path lie on the first (resp., the second) sheet 
  of $\overline{\Sigma}$.}
  \label{fig:weber-path}
 \end{figure}
\end{exc}

\subsection{Stokes graph and Borel summability}

To give a rigorous analytic realization of WKB solutions, 
we use the so-called {\em Borel summation method} 
and \'Ecalle's {\em resurgent analysis}.
See \cite{Ecalle, Costin, Sauzin} for details. 

\subsubsection{Brief review of Borel summation method}
\label{subsubsec:borel-summation-method}

Suppose we have a compact set $K$ such as in Proposition \ref{prop:gevrey-1}. 
The {\em Borel transform} of the WKB solution \eqref{eq:wkb-series}
is defined to be its term-wise inverse Laplace transform\footnote{
Here we shift the lower endpoint of the Laplace integral to absorb the exponential factor as follows:
\[
{\rm e}^{\pm S/\hbar} \hbar^{\alpha} = 
\int^{\infty}_{\mp S} {\rm e}^{- \zeta/\hbar} \frac{(\zeta \pm S)^{\alpha - 1}}{\Gamma(\alpha)} \, {\rm d}\zeta
\qquad ({\rm Re} \, \alpha > 0).
\]
} as
\begin{equation}
{\mathcal B}\psi_{\pm}(x, \zeta) = 
\psi_{\pm, B}(x,\zeta) = \sum_{m \ge 0} \frac{\psi_{\pm,m}(x)}{\Gamma(m+\frac{1}{2})} (\zeta \pm S(x))^{m - \frac{1}{2}}.
\end{equation}
Thanks to the estimate \eqref{eq:gevrey-1}, for any $x \in K$, 
the above series $\psi_{\pm, B}(x,\zeta)$ converges and defines a germ of (multi-valued) function 
on a punctured neighborhood of $\zeta = \mp S(x)$. 
If it has an analytic continuation with respect to $\zeta$
to a domain containing the half line $\mp S(x) + {\mathbb R}_{\ge 0}$, 
and growths at most exponentially when $|\zeta| \to +\infty$, 
then the Laplace transform 
\begin{equation} \label{eq:BS}
{\mathcal S} \psi_{\pm}(x,\hbar) = 
\int_{\mp S(x)}^{\infty} {\rm e}^{-\zeta/\hbar} 
\psi_{\pm, B}(x,\zeta) \, {\rm d}\zeta
\end{equation}
converges for $0 < \hbar \ll 1$. 
We say that the WKB solution $\psi_\pm$ is {\em Borel summable} (as a series of $\hbar$) 
on a domain $D$ if the Laplace integral \eqref{eq:BS} converges for any compact subset $K \subset D$.  
We call the resulting function \eqref{eq:BS}, 
defined on the domain $D$, the {\em Borel sum} of $\psi_\pm$.
One of the important properties of the Borel sum is that it 
recovers the original WKB solution as an asymptotic expansion (for any fixed $x$):
\[
{\mathcal S} \psi_{\pm}(x,\hbar) \sim \psi_{\pm}(x,\hbar), \quad \hbar \to +0.
\]

\smallskip
\begin{exc} \label{exc:airy-borel}
Prove that the Borel transform $\psi^{\rm Airy}_{\pm, B}(x,\zeta)$ of the WKB solution \eqref{eq:wkb-airy} 
of the Airy equation can be explicitly written as follows:
\begin{equation} \label{eq:airy-borel-2f1}
\begin{cases}
\displaystyle 
\psi^{\rm Airy}_{+, B}(x,\zeta) = C \, x^{-1} w^{- \frac{1}{2}} 
{}_{2}F_{1}\Bigl( \frac{1}{6}, \frac{5}{6}, \frac{1}{2}; w \Bigr), 
\\[+1.0em]
\displaystyle
\psi^{\rm Airy}_{-, B}(x,\zeta) = C \,  x^{-1} (w-1)^{- \frac{1}{2}} 
{}_{2}F_{1}\Bigl( \frac{1}{6}, \frac{5}{6}, \frac{1}{2}; 1 - w \Bigr).
\end{cases}
\end{equation} 
Here,  
\begin{equation}
C = \frac{\sqrt{3}}{2\sqrt{\pi}}, \qquad  
w = \frac{\zeta + S(x)}{2 S(x)}, \qquad 
S(x) = \frac{2}{3}x^{3/2},
\end{equation} 
and ${}_{2}F_{1}$ is the {\em Gauss hypergeometric function} defined 
for $\alpha, \beta \in {\mathbb C}$ and $ \gamma \in {\mathbb C} \setminus {\mathbb Z}_{\le 0} $ by
\begin{equation}
{}_{2}F_{1}( \alpha, \beta, \gamma; w) = \sum_{n \ge 0} \frac{(\alpha)_n (\beta)_n}{(\gamma)_n n!} w^n,  
\end{equation}
where $(a)_n$ denotes the Pochhammer symbol:
\begin{equation}
(a)_n = 
\begin{cases} 
1 & n = 0, \\
a(a+1) \cdots (a+n-1) & n \ge 1.
\end{cases}
\end{equation}
\end{exc}
It is well known that the hypergeometric function 
${}_{2}F_{1}(\alpha, \beta, \gamma; w)$ 
satisfies the following linear ODE, 
known as the {\em Gauss hypergeometric differential equation}:
\begin{equation}
\left( \frac{{\rm d}^2}{{\rm d}w^2} + \frac{\gamma - (\alpha + \beta + 1)w}{w(1-w) } \frac{\rm d}{{\rm d}w} 
- \frac{\alpha \beta}{w(1-w) } \right) F = 0.
\end{equation}
The equation possesses regular singularity at $w=0, 1, \infty$.
Therefore, ${}_{2}F_{1}(\alpha, \beta, \gamma; w)$ can be analytically continued along 
any path that avoids these singularities.
The explicit expression \eqref{eq:airy-borel-2f1} shows that 
the Borel transform $\psi^{\rm Airy}_{\pm,B}(x,\zeta)$, which is originally defined as a convergent series defined near 
$\zeta = \mp S(x)$, has a singularity at $\zeta = \pm S(x)$. 
Therefore, the WKB solution $\psi^{\rm Airy}_\pm$ are Borel summable 
on each connected component of the compliment of the locus defined by ${\rm Im} \, S(x) = 0$. 
Such Borel summability conditions are more generally provided in 
the next subsection using the concept of a Stokes graph.

\subsubsection{Stokes graphs}

Since the coefficients of the WKB solution are functions of $x$, 
its Borel summability does depend on the position of $x$. 
The Stokes graph introduced here provides a criterion for determining 
the Borel summability of the WKB solution. 

\begin{dfn}
The {\em Stokes graph} of the Schr\"odinger-type ODE \eqref{eq:sch} 
is a graph 
whose
\begin{itemize}
\item 
vertices are give by zeros and poles of $\phi$, and
\item 
edges are given by {\em Stokes curves}, which are real 1-dimensional curves 
emanating from a turning point $v$ defined by 
\begin{equation}
{\rm Im}\, \int^{x}_{v} 
\sqrt{Q(x')} \, {\rm d}x' 
= 0.
\end{equation}
\end{itemize}
\end{dfn}

  \begin{figure}[t]
\begin{minipage}{0.3\hsize}
  \begin{center} 
  \includegraphics[width=35mm]
  {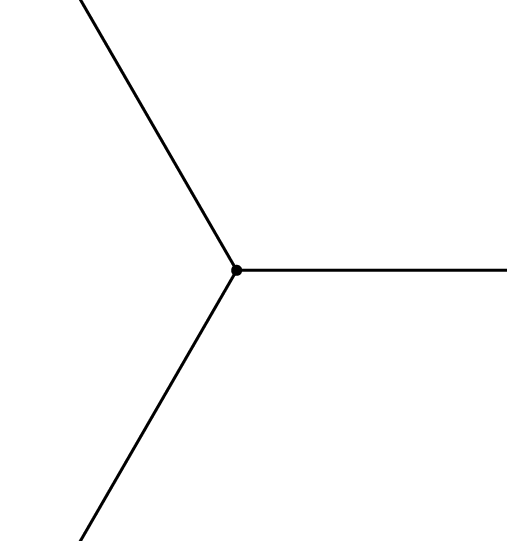} \\[+.5em]
  {\small (a) $Q(x) = x$} 
  \end{center}
\end{minipage} \hspace{+.5em}
\begin{minipage}{0.3\hsize}
  \begin{center}
  \includegraphics[width=35mm]
  {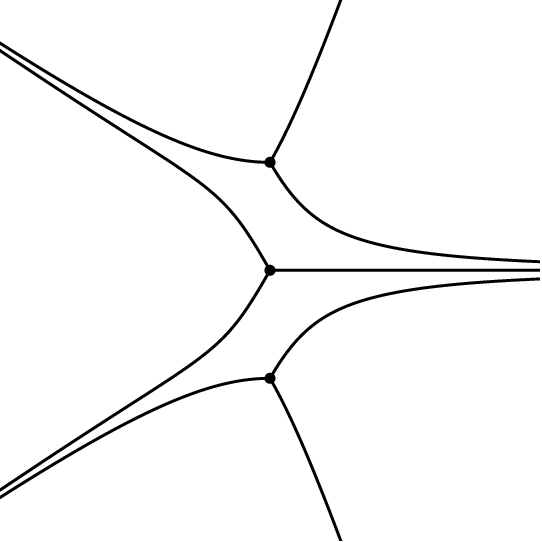} \\[+1.2em]
  {\small (b) $Q(x) = x^3+x$} 
  \end{center}
\end{minipage}  \hspace{+.5em}
\begin{minipage}{0.3\hsize}
  \begin{center} 
  {\color{white} oooooo}
  \includegraphics[width=30mm]
  {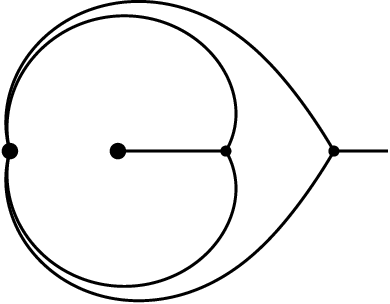} \\[+4.8em]
  {\small (c) $Q(x) = \displaystyle \frac{(x-1)(x-2)}
  {x^{2}(x+1)^{2}}$} 
  \end{center}
 \end{minipage}  
 
 \vspace{+1.em}
 
  \begin{minipage}{0.3\hsize}
  \begin{center}
  \includegraphics[width=40mm]{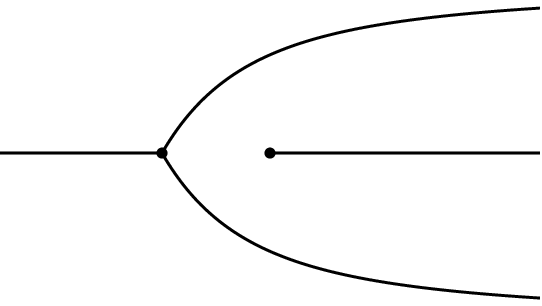} \\[+5.0em]
  \small{(${\rm d}$) $Q(x) = \dfrac{x+1}{x}$}
  \end{center}
  \end{minipage}  \hspace{+.5em}
  \begin{minipage}{0.3\hsize}
  \begin{center}
  \includegraphics[width=35mm]{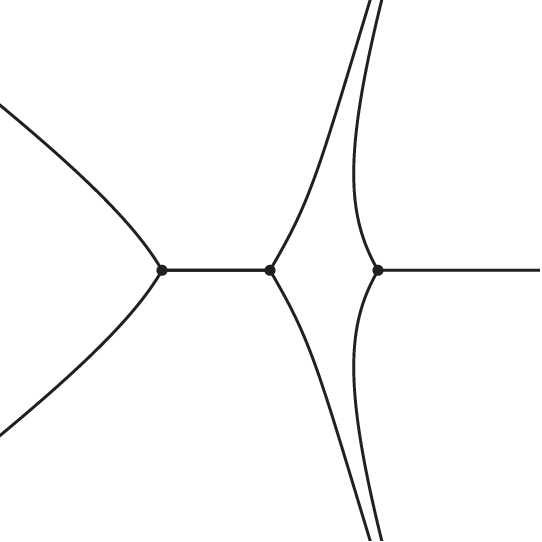} \\[+.8em]
  {\small (${\rm e}$) $Q(x) = x^3-x$}
  \end{center}
  \end{minipage} 
  \begin{minipage}{0.3\hsize}
  \begin{center}
  {\color{white} oooo}
  \includegraphics[width=31mm]{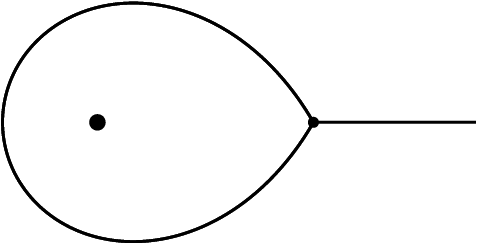} \\[+6.5em]
  {\small (${\rm f}$) $Q(x) = \dfrac{x-2}{x^2}$}
  \end{center}
  \end{minipage}  \hspace{+.5em}

  \caption{Example of Stokes graphs.}
  \label{fig:Stokes-example}
  \end{figure}

The Stokes curves are nothing but the (horizontal) {\em trajectories} of
the meromorphic quadratic differential $\phi$.  
The Stokes graph is identical to the {\em spectral network} 
for a class $S$ theory specified by $\phi$ (see \cite{GMN09, GMN12}).
Figure \ref{fig:Stokes-example} shows some examples of Stokes graphs.
In general, Stokes curves satisfy the following properties 
(see \cite{Str, BS} for details): 
\begin{itemize}
\item 
Three Stokes curves emanate from a simple zero of $\phi$, while 
one Stokes curve emanates from a simple pole of $\phi$.
\item 
There are $(m-3)$ asymptotic directions of Stokes curves 
around a pole of $\phi$ with order $m \ge 3$. 
\item 
Local structure of Stokes curves around a double pole of $\phi$ 
depends on the argument of the residue of $\sqrt{\phi}$ at the pole. 
\end{itemize}

The interior of each face of the Stokes graph is called a {\em Stokes region}. 
In (e) and (f), we can observe that there exists a Stokes curve connecting (possibly the same) turning points. 
Such a Stokes curve is called a {\em saddle connection} (or a {\em Stokes segment}). 

\begin{rem}
Let us take $\phi(x) = (1+{\rm i}) \, {\rm d}x^{2}/(x^4-1)$. 
One Stokes curve emanates from each of the four simple poles of $\phi$. 
However, since the quadratic differential $\phi$ has no pole of order $\ge 2$
(therefore it does not satisfy Assumption \ref{ass:part1}), 
these Stokes curves do not have another end and form recurrent trajectories (c.f., \cite{thabet}). 
Under Assumption \ref{ass:part1}, it is known that no recurrent trajectory appears 
if there is no saddle connection (see \cite[Lemma 3.1]{BS}). 
\end{rem}


\subsubsection{Borel summability and idea of proof} 
\label{subsubsec:koike-proof}

Now we can give the criterion of the Borel summability of the WKB solutions as follows.  

\begin{thm} \label{thm:summablity}
Assume that the Stokes graph of \eqref{eq:sch} does not contain any saddle connection. 
Then, the WKB solution $\psi_\pm$ 
is Borel summable on each Stokes region. 
Also, for a sufficiently small $\hbar > 0$, 
the Borel sum ${\mathcal S}\psi_{\pm}$ gives a holomorphic solution of 
the Schr\"odinger-type ODE \eqref{eq:sch} on the Stokes region. 
\end{thm}

Theorem \ref{thm:summablity} was proved by several works including
\cite{DLS, Koike-Schafke, Nemes, Nikolaev}. 
Below, we will very roughly explain why the trajectory of the quadratic differential $\phi$
controls the Borel summability, following the idea due to Koike--Sch\"afke. 
See also \cite[\S 3]{Takei-review}. 

First, we will investigate the Borel transform of an auxiliary series 
\begin{equation}
T(x,\hbar) = \sum_{m \ge 1} \hbar^{m} P_m(x) 
~~ \left( =  P(x,\hbar) - \hbar^{-1} P_{-1}(x) - P_0(x)  \right).
\end{equation}
The Riccati equation for $P$ implies 
\begin{equation} \label{eq:ode-for-T}
2 \sqrt{Q(x)} \, \left( \hbar^{-1} T(x,\hbar) - P_{1}(x) \right) + \frac{{\rm d}T}{{\rm d}x}(x,\hbar) = 
- 2 P_0(x) T(x,\hbar) - T(x,\hbar)^2. 
\end{equation}
Now, let us change the local coordinate, known as the Liouville transformation, 
from $x$ to $z$ defined by 
\begin{equation} \label{eq:Liouville}
z = z(x) = \int^{x}_{x_\ast} 
\sqrt{Q(x')} \, {\rm d}x'
\end{equation} 
with a certain reference point $x_\ast$, and 
take the the Borel transform of the both sides.  
Since the multiplication by $\hbar^{-1}$ is translated into
the derivative $\partial_\zeta$ for the Borel transform, we have
\begin{equation}
\left( \frac{\partial}{\partial \zeta} + \frac{1}{2} \frac{\partial}{\partial z}  \right)  T_B(z, \zeta) 
= A_1(z) T_B (z, \zeta) + A_2(z) \frac{\partial}{\partial \zeta}  T_B \ast T_B(z, \zeta), 
\end{equation}
where\footnote{
\label{footnote:summability}
If we consider the case where the potential $Q$ depends on $\hbar$, 
as mentioned in Remark \ref{rem:hbar-correction}, 
its Borel transform $Q_B$ is added to $A_3$ in a simple manner. 
Therefore, the following discussion is easily generalized to this case, 
provided that the potential $Q$ is Borel summable with respect to $\hbar$.
} 
\begin{equation} \label{eq:Ai-coeffs}
A_1\bigl( z(x) \bigr) = - \frac{P_0(x)}{\sqrt{Q(x)}}, \quad
A_2\bigl( z(x) \bigr) = - \frac{1}{2\sqrt{Q(x)}}, 
\end{equation}
and $\ast$ is the convolution product with respect to $\zeta$ defined by
\begin{equation}
T_B \ast T_B(z, \zeta) = \int^{\zeta}_{0} T_B(z, \sigma) T_B(z, \zeta - \sigma) \, {\rm d}\sigma.
\end{equation}  
Then, we can derive the following integro-differential equation for $T_B$:
\begin{align} 
T_B(z,\zeta) & = A_0\left(z - \frac{\zeta}{2} \right)
+ \int^{\zeta}_{0} A_1\left(z - \frac{\zeta-t}{2} \right)
T_B \left(z - \frac{\zeta-t}{2}, t \right) \, dt 
\notag \\[+.7em] 
& + \int^{\zeta}_{0} A_2\left(z - \frac{\zeta-t}{2} \right)
\frac{\partial}{\partial \zeta} T_B \ast T_B \left(z - \frac{\zeta-t}{2}, t \right) \, {\rm d}t 
\label{eq:integral-eq}
\end{align}
where 
\begin{equation} \label{eq:A0-coeffs}
A_0(z(x)) = T_B(z(x), 0) = P_1(x).
\end{equation}

Now, let us take a reference point $x_\ast$, which is neither a turning point nor a singular point,
and ask if $T_B (z(x_\ast), \zeta)$ can be analytically continued along the positive real axis on $\zeta$-plane.
Koike--Sch\"afke's approach is to show the existence of the analytic continuation 
by constructing a sequence of successive approximations for equation \eqref{eq:integral-eq} 
and performing various estimates to prove its convergence. 
We omit the construction of the sequence of successive approximations; 
however, the boundedness of the coefficients $A_i\left(z - \frac{\zeta-t}{2} \right)$, 
appearing in \eqref{eq:integral-eq}, 
is essential for estimating the successive approximations when 
$\zeta$ moves in the positive direction along the real axis.

As $\zeta$ moves in the positive direction along the real axis 
(with $t$ being satisfying $0 \le t \le \zeta$), 
the point $z - (\zeta - t)/2$, which is appearing in \eqref{eq:integral-eq}, 
moves in the direction where its real part decreases while keeping its imaginary part. 
We should also notice that the pullback of the lines parallel to the real axis in the $z$-plane 
via the Liouville transformation \eqref{eq:Liouville} corresponds precisely to the trajectories of $\phi$. 
Therefore, if we denote by $p$ the point reached by the negative part of the trajectory passing through $x_\ast$ 
(i.e., the set of points $x$ on the trajectory satisfying ${\int^{x}_{x_\ast}} \sqrt{Q(x')}\,{\rm d}x' < 0$), 
the boundedness of $A_i(z(x))$ in the vicinity of $p$ becomes the issue for estimating the successive approximations. 
(Note that the negative part of trajectory depends on the choice of the branch of $\sqrt{Q(x)}$.)
There are two possibilities for such $p$:

\begin{itemize}
\item
The case where $p$ is a turning point. This case happens when $x_\ast$ lies on a Stokes curve 
emanating from $p$, and satisfies $\int^{x_\ast}_{p} \sqrt{Q(x')}\, {\rm d}x' > 0$. 
Since $A_i(z(x))$, given in \eqref{eq:Ai-coeffs} and \eqref{eq:A0-coeffs}, 
is singular at turning points, 
the boundedness of the coefficient in \eqref{eq:integral-eq}, 
which is necessary for evaluating the successive approximation sequence, 
cannot be guaranteed in this case.
Consequently, the existence of the analytic continuation of 
$T_B(z, \zeta)$  cannot be established.

\item
The case where $p$ is a pole of the quadratic differential $\phi$ 
of order greater than or equal to $2$. 
This situation arises when 
$x_\ast$ is either contained within a certain Stokes region 
or located on the negative part of a certain Stokes curve.
To simplify the discussion, we assume 
that the pole order at $p$ is greater than $2$.
(In fact, we can relax this assumption by a slight modification of the method.)
Then, we can verify that $A_i(z(x))$ is bounded in a neighborhood of $p$ (c.f., Exercise \ref{exc:pole-order}),
and the sequence of successive approximation can be constructed, 
and an appropriate estimate of it can be obtained.
Consequently, we can establish the existence of the analytic continuation of $T_B(z, \zeta)$, 
the growth rate of $T_B(z, \zeta)$ as $|\zeta| \to +\infty$ is shown to be at most exponential.
This result confirms the Borel summability of $T(x,\hbar)$ at $x = x_\ast$. 
 
\end{itemize}

Thus we have constructed the Borel sum of $T(x,\hbar)$ for any $x$ in the complement of the Stokes graph. 
The details are omitted in this note, but the above estimates can be carried out uniformly with respect to $x$.
That is, it is also shown that the constructed Borel sum of $T(x,\hbar)$ 
indeed provides a solution to the differential equation \eqref{eq:ode-for-T}.

\begin{figure}[t]
 \begin{minipage}{0.5\hsize}
  \begin{center}
   \includegraphics[width=50mm]{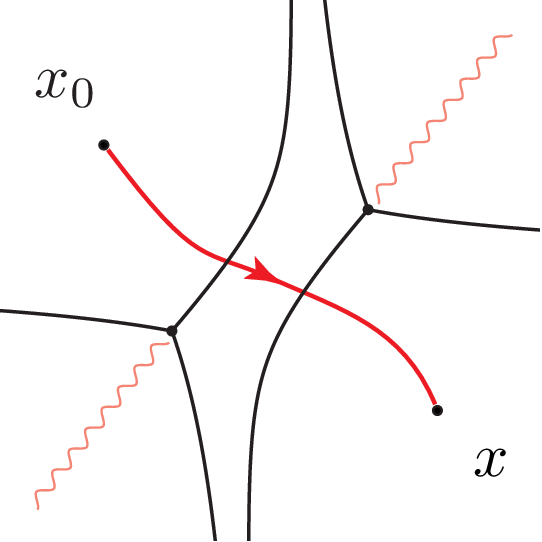} 
  \end{center}
 \end{minipage}
  =  %
  \begin{minipage}{0.5\hsize}
  \begin{center}
   \includegraphics[width=50mm]{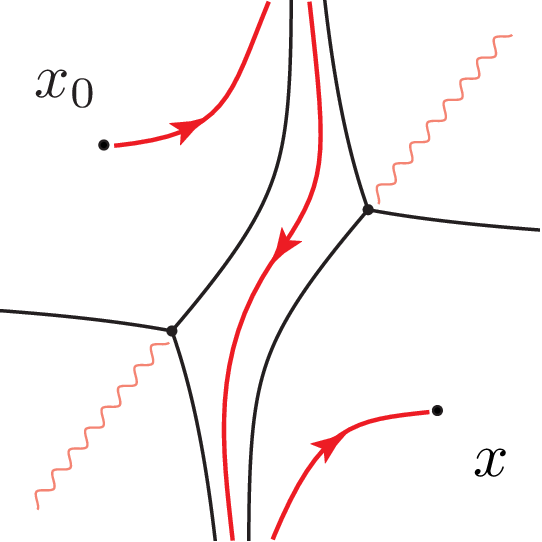} 
  \end{center}
 \end{minipage}
   \caption{When there are no saddle connections in the Stokes graph, 
   a path (in red) intersecting with Stokes curves, as shown in the left figure, 
   can be decomposed into several paths that do not intersect with any Stokes curves, 
   as shown in the right figure.}
  \label{fig:koike-trick}
\end{figure}

Next, let us consider the Borel summability of $\int^{x}_{x_0} T(x', \hbar) \, {\rm d}x'$, 
a core part of the WKB solution. 
If it is Borel summable, then the Borel summability of the WKB solution 
follows from \cite[Theorem 13.3]{Sauzin}.
The point that needs to be considered here is the possibility that the integration path
from $x_0$ to $x$ may intersect with several Stokes curves. 
From the previous discussion, it is not guaranteed that 
the integrand $T$ is Borel summable at these intersection points.
However, if the integration path can be decomposed into 
a number of paths which are contained in Stokes regions, 
where the endpoints of each path in the decomposition are chosen at poles of $\phi$
(see the Figure \ref{fig:koike-trick}). 
Then, we can verify that the $\int^{x}_{x_0} T(x', \hbar) \, {\rm d}x'$ becomes Borel summable
since the integrant $T$ is Borel summable at any point on the modified path. 
This is what we call ``{\em Koike's trick}''.  

Such a deformation is impossible if the original path 
has nontrivial intersection with saddle connections\footnote{
We can also relax the assumption in Theorem \ref{thm:summablity}. 
That is, even if there exist saddle connections, 
the WKB solution is Borel summable if the path of integration in \eqref{eq:wkb-sol} 
from $x_0$ to $x$ never intersect with saddle connections. 
\label{footnote:saddle}
}. 
This is the reason why the saddle connection gives an obstruction for the Borel summability of WKB solutions
(see also Remark \ref{rem:Stokes-Weber-irreg} below).

%
%
%

\subsection{Connection formulas on Stokes curves}
\label{sec:conn-formula}

In general, the exact solutions of the Schr\"odinger-type ODE \eqref{eq:sch}
constructed by the Borel summation method differ across Stokes regions. 
In this section, we will see the connection formulas that describe 
how these Borel sums are related.

 \begin{figure}[b]
  \begin{center} 
  \includegraphics[width=60mm]{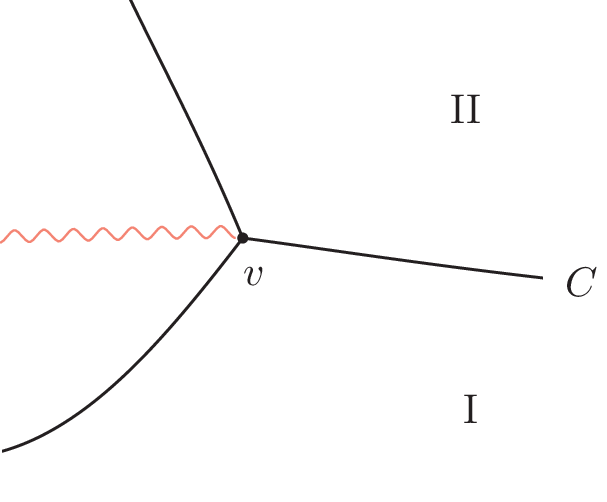} 
  \end{center}  
  \caption{Stokes curve $C$ and adjacent Stokes regions ${\rm I}$, ${\rm II}$.}
  \label{fig:voros1}
 \end{figure}

\subsubsection{Voros connection formula around a simple turning point}
\label{subsec:voros-formula}

Here, we will see the Voros connection formula. 
First, let us describe the setup. 

\begin{ass} ~ 
\begin{itemize}
\item[({\rm i})] Let $v$ is a simple zero of $\phi$, and 
$C$ be a Stokes curve emanating from $v$ 
which forms a common boundary of two adjacent Stokes regions ${\rm I}$ and ${\rm II}$.
From $v$, region ${\rm II}$ appears next to region ${\rm I}$ in the counter-clockwise direction
(see Figure \ref{fig:voros1}).

\item[({\rm ii})]
The Stokes graph of \eqref{eq:sch} does not contain any saddle connection.

\end{itemize}
\end{ass}

In order to write down the connection formulas, 
we must fix a normalization of the WKB solution. 
Here, we adopt the normalization that takes 
the turning point $v$ at the lower endpoint
in the integral in \eqref{eq:wkb-sol}:
\begin{equation} \label{eq:WKB-TP}
\psi_{\pm}(x,\hbar) = 
\frac{1}{\sqrt{ P_{\rm odd}(x,\hbar)}}
\exp \left( \pm \int^x_{v} P_{\rm odd}(x,\hbar) \, {\rm d}x \right).
\end{equation}
Here we note that, 
although the coefficients of $P_{\rm odd}$ has singularity at $v$, 
the integral with the endpoint $v$ can be defined by means of a contour integration 
(see \cite[\S 2.1]{KT98} for details).
We also denote by $\Psi_{\pm}^{\rm J}$ the Borel sum of $\psi_{\pm}$
defined in the Stokes region ${\rm J} = {\rm I}, {\rm II}$. 
Then we have the following, which we call 
the {\em Voros connection formula} around a simple turning point.

\begin{thm}[{\cite{Voros83}, \cite{Silverstone}, \cite[Theorem 2.23]{KT98}, \cite{KK12}}] \label{thm:voros-conn}
Under the above setting, the analytic continuation of 
$\Psi_{\pm}^{\rm I}$ to the Stokes region ${\rm II}$ across the Stokes curve $C$ 
is described as
\[
(\Psi_{+}^{\rm I}, \Psi_{-}^{\rm I}) 
= 
(\Psi_{+}^{\rm II}, \Psi_{-}^{\rm II}) \cdot S, 
\]
\begin{equation} \label{eq:Stokes-Voros}
S = \begin{cases}
\begin{pmatrix} 1 & 0 \\ {\rm i} & 1 \end{pmatrix} & ~~
\text{if $\displaystyle \int^x_{v} 
\sqrt{Q(x')} \, {\rm d}x' > 0$ on $C$}, 
\\[+1.5em]
\begin{pmatrix} 1 & {\rm i} \\ 0 & 1 \end{pmatrix} & ~~
\text{if $\displaystyle \int^x_{v} 
\sqrt{Q(x')} \, {\rm d}x' < 0$ on $C$}. 
\end{cases}
\end{equation}
\end{thm}

\begin{figure}[h]
  \begin{center} 
  \includegraphics[width=40mm]{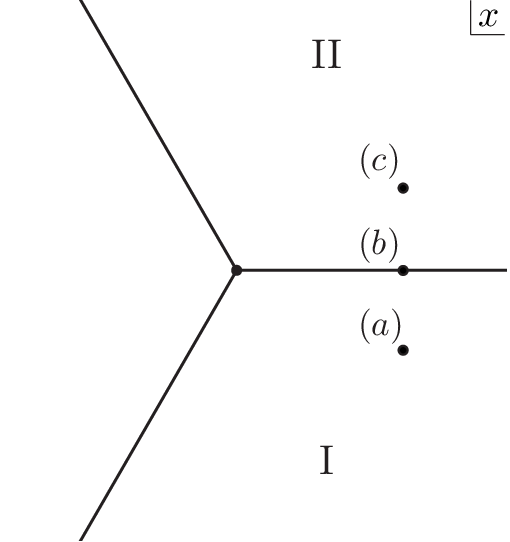}  
  \end{center}  
\centering
\begin{minipage}[b]{0.325\columnwidth}
    \centering
    \includegraphics[width=0.98\columnwidth]{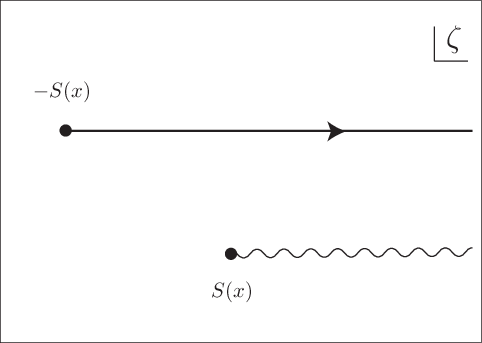} \\ (a) 
\end{minipage} 
\hspace{-.5em}
\begin{minipage}[b]{0.325\columnwidth}
    \centering
    \includegraphics[width=0.98\columnwidth]{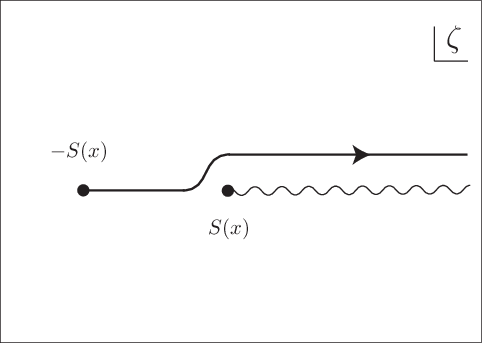} \\ (b)
\end{minipage}
\hspace{-.5em}
\begin{minipage}[b]{0.325\columnwidth}
    \centering
    \includegraphics[width=0.98\columnwidth]{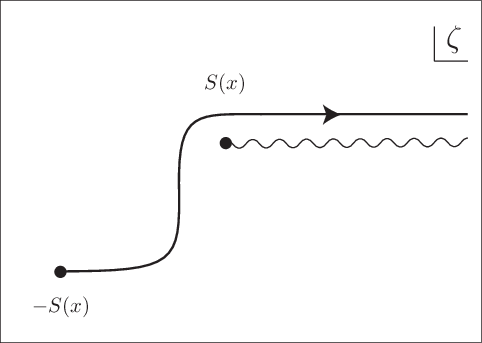} \\ (c)
\end{minipage}
\caption{The upper row figure depicts the Stokes graph of the Airy equation, 
and the figures (a)--(c) in the lower row represent the Borel planes 
corresponding to the positions (a)--(c) of $x$ in the upper row figure. 
The half-line emanating from $-S(x)$ in (a) represents the path of Laplace integration 
that defines the Borel sum $\Psi_{+}^{\rm Airy, \, I}$, 
while the integration paths providing the analytic continuation to region ${\rm II}$ 
with respect to $x$ are shown in Figures (c). The wiggly line represents 
a branch cut to describe the multivaluedness of $\psi_{+,B}^{\rm Airy}$.}
\label{fig:airy-borel-plane}
\end{figure}

Let us explain how the connection formula \eqref{eq:Stokes-Voros} 
is derive in the case of the Airy equation \eqref{eq:airy-eq}, 
following the discussion of \cite[\S 2.2]{KT98}. 
Here we focus on the connection formula on the positive real axis on $x$-plane, 
which is a Stokes curve of the Airy equation as we have seen in Figure \ref{fig:Stokes-example} (a); 
see also Figure \ref{fig:airy-borel-plane} below.
We take a principal branch of $\sqrt{Q(x)} = \sqrt{x}$ on the Stokes curve, 
so that $S(x) = \int^x_{0} \sqrt{Q(x')} \, {\rm d}x' = \frac{2}{3} x^{3/2} > 0$ holds there.
Let us focus on the connection formula of $\psi^{\rm Airy}_{+}(x,\hbar)$; 
for the purpose, let us look at the singularity structure of the Borel transform 
$\psi^{\rm Airy}_{+,B}(x,\zeta)$ on the Borel plane (i.e., $\zeta$-plane).

\begin{figure}[h]
 \begin{minipage}{0.5\hsize}
  \begin{center}
   \includegraphics[width=60mm]{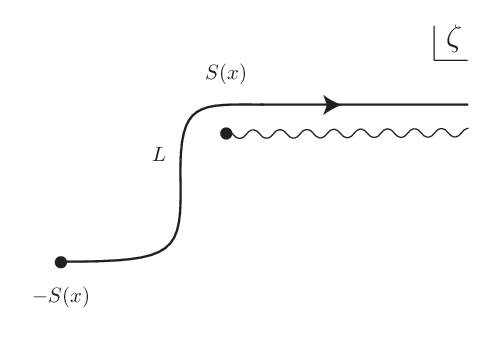} \\{(i)}
  \end{center}
 \end{minipage}
 \hspace{-1.em} = \hspace{-1.em} %
  \begin{minipage}{0.5\hsize}
  \begin{center}
   \includegraphics[width=60mm]{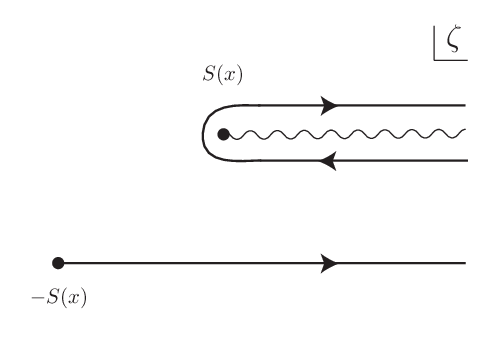} \\{(ii)}
   \end{center}
 \end{minipage}
    \caption{(i) The integration path $L$ that provides the analytic continuation of 
    $\Psi_{+}^{\rm Airy,\,I}$ to region II. 
    (ii) The integration path obtained as a continuous deformation of $L$. 
    The integral along the half-line emanating from $-S(x)$ corresponds to 
    the first term in equation \eqref{eq:AC-to-region-II-modified} 
    (i.e., $\Psi_{+}^{\rm Airy,\,II}$), 
    while the integral around the branch cut originating from $S(x)$ corresponds 
    to the second term in equation \eqref{eq:AC-to-region-II-modified}.}
    \label{fig:deformation-Laplace-contour}
\end{figure}

We have already seen the location of the singularities of $\psi^{\rm Airy}_{+, B}$ 
at the end of \S \ref{subsubsec:borel-summation-method}.
The half-line extending to the right from $\zeta = -S(x)$ 
in Figure \ref{fig:airy-borel-plane} (a)
represents the Laplace integration path that defines the Borel sum $\Psi_{+}^{\rm Airy,\,I}$ defined on region I.
As $x$ moves continuously from region I to region II across the Stokes curve, 
the Borel singularity $\zeta = S(x)$ crosses the Laplace integration path. 
Accordingly, to describe the analytic continuation of $\Psi_{+}^{\rm Airy,\,I}$ to region II, 
it is necessary to continuously deform the Laplace integration path
as is shown in \ref{fig:airy-borel-plane} (a)--(c). 
The resulting analytic continuation is given by the integral 
\begin{equation} \label{eq:AC-to-region-II}
\int_{L} e^{- \zeta/\hbar} \psi^{\rm Airy}_{+,B}(x,\zeta) \, {\rm d}\zeta
\end{equation}
along the path $L$ shown in Figure \ref{fig:deformation-Laplace-contour} (i).
Furthermore, by continuously deforming the integration path as shown in 
Figure \ref{fig:deformation-Laplace-contour} (ii), 
it can be seen that \eqref{eq:AC-to-region-II} matches 
\begin{equation} \label{eq:AC-to-region-II-modified}
\int_{-S(x)}^{\infty} e^{- \zeta/\hbar} \psi^{\rm Airy}_{+,B}(x,\zeta) \, {\rm d}\zeta
+ \int_{+S(x)}^{\infty} e^{- \zeta/\hbar} \Bigl( \dot{\Delta}_{\zeta = + S(x)} 
\psi^{\rm Airy}_{+,B}(x,\zeta) \Bigr)\, {\rm d}\zeta,
\end{equation}
where $\dot{\Delta}_{\zeta = + S(x)} \psi^{\rm Airy}_{+,B}(x,\zeta)$
is the discontinuity\footnote{
In the resurgence theory, the symbol $\dot{\Delta}_{\omega}$ is used for the 
so-called {\em (dotted) alien derivative} associated with a Borel singularity $\omega$; 
see \cite{Ecalle, Sauzin} for the definition. 
For formal power series with half-integer power such as \eqref{eq:wkb-series}, 
the dotted alien derivative is translated into the discontinuity (c.f., \cite[Appendix B]{AIT19}).
Therefore we have used the same notation. 
} 
of $\psi^{\rm Airy}_{+,B}(x,\zeta)$ at $\zeta = + S(x)$, 
which is defined as the difference of the boundary values from above and below the positively oriented 
straight line emanating from $\zeta = + S(x)$. 
Then, the connection formula  
\begin{align}
{}_{2}F_{1}\left( \alpha, \beta, \gamma; w \right) 
& =
\frac{\Gamma(\gamma)\Gamma(\alpha + \beta - \gamma)}{\Gamma(\alpha) \, \Gamma(\beta)} 
(1-w)^{\gamma - \alpha - \beta}  {}_{2}F_{1} \left( \gamma - \alpha ,  \gamma - \beta, \gamma - \alpha - \beta + 1 ; 1- w \right) \notag \\ 
& \quad + \frac{\Gamma(\gamma) \Gamma(\gamma - \alpha - \beta)}{\Gamma(\gamma - \alpha) \, \Gamma(\gamma - \beta)}  
{}_{2}F_{1} \left( \alpha, \beta,  \alpha + \beta - \gamma + 1,; 1- w \right), 
\label{eq:gauss-connection-formula}
\end{align}
for the Gauss hypergeometric functions between $w = 0$ to $1$ 
which is valid if $\gamma, \gamma - \alpha -\gamma \notin {\mathbb Z}$
(e.g., \cite[Corollary 2.3.3]{ASR1999}), 
and the transformation property
\begin{equation}
{}_{2}F_{1}\left( \alpha, \beta, \gamma; w \right) 
= (1 - w)^{\gamma - \alpha - \beta} {}_{2}F_{1}\left(\gamma - \alpha, \gamma - \beta, \gamma; w \right) 
\end{equation}
which is valid if $\gamma, \gamma - \beta \notin {\mathbb Z}$ (e.g., \cite[Theorem 2.2.5]{ASR1999}) imply
\begin{equation}
\dot{\Delta}_{\zeta = + S(x)} \psi^{\rm Airy}_{+,B}(x,\zeta) 
= {\rm i} \, \psi^{\rm Airy}_{-,B}(x,\zeta). 
\end{equation}
Therefore, we have confirmed that \eqref{eq:AC-to-region-II-modified} 
precisely agrees with $\Psi_{+}^{\rm Airy,\,II} + {\rm i} \, \Psi_{-}^{\rm Airy,\,II}$.

On the other hand, we can also verify that the WKB solution 
$\psi^{\rm Airy}_{-}(x,\hbar)$ is Borel summable even when 
$x$ lies on the Stokes curve in question. 
This is evident from the fact that the Borel transform $\psi^{\rm Airy}_{-,B}(x,\zeta)$ 
has no other singularities on $\zeta$-plane along the Laplace integration path, 
and it is also a consequence of the general theory explained in \S \ref{subsubsec:koike-proof}.
Thus, the Borel sum does not exhibit discontinuity on the Stokes curve, 
and we have $\Psi^{\rm Airy,\,I}_{-} = \Psi^{\rm Airy,\,II}_{-}$.
Thus we have proved the connection formula \eqref{eq:Stokes-Voros} for the Airy equation. 

The connection formula for the WKB solutions of 
general Schr\"odinger-type ODEs are obtained by 
the {\em exact WKB theoretic reduction} to the Airy equation. 
See \cite[\S 2.3]{KT98} and \cite{KK12} for details.
Since it is derived through the analysis of Borel singularity, 
the formula \eqref{eq:Stokes-Voros} can also be understood as 
explicitly describing the {\em Stokes phenomenon} for the divergent WKB solution 
(and this is the origin of the name  ``Stokes'' graph).

\smallskip
\begin{exc} 
Show that the Borel sum of the WKB solutions are single-valued around
any simple zero of $\phi$.
\end{exc}

\subsubsection{Relation to the path-lifting rule of Gaiotto--Moore--Neitzke}
\label{subsec:path-lifting}

\begin{figure}[t]
  \begin{center} 
  \includegraphics[width=55mm]{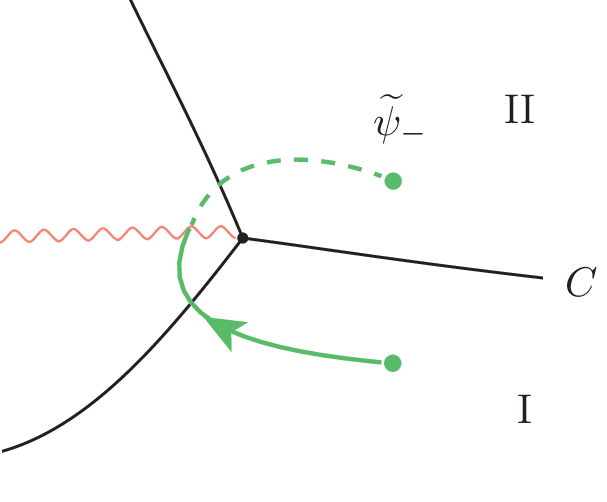} \\[-3.em] {\color{white} ---}
  \end{center}  
  \caption{The detoured path.}
  \label{fig:voros2}
 \end{figure}

Here we give an alternative description of the Voros connection formula. 
Assume that we are in the same situation as the previous subsection, 
and further assume that 
$\int^{x}_{v} \sqrt{Q(x')} \, {\rm d}x' > 0$ 
holds on $C$. 
Then, the formula \eqref{eq:Stokes-Voros} for $\psi_+$ is equivalent to 
\begin{equation} \label{eq:path-lifting-formula}
\Psi_{+}^{\rm I} = \Psi_{+}^{\rm II} + \widetilde{\Psi}_{-}^{\rm II}, 
\end{equation}
where the second term is the Borel sum of the WKB solution $\widetilde{\psi}_-$ 
obtained as the term-wise analytic continuation of $\psi_+$ along the ``detoured path'' 
depicted in Figure \ref{fig:voros2}.
The formula \eqref{eq:path-lifting-formula} is actually stated 
in this form in Voros' original paper; see \cite[eq.\,(6.19)]{Voros83}.
Since the path crosses the branch cut and terminates at the different sheet, 
we put the subscript $-$ for the resulting WKB solution.  
The description is equivalent to the {\em path-lifting rule}, 
which was named in their study of {\em (non-)abelianization} by Gaiotto--Moore--Neitzke \cite{GMN12}. 
We also note that the above path-lifting rule \eqref{eq:path-lifting-formula} 
is valid for any WKB solutions, including those which may not be normalized 
at the turning point, such as \eqref{eq:WKB-TP}.


\smallskip

\bigskip
\subsubsection{Koike connection formula around a simple pole} \label{subsec:koike-formula}

It was shown by Koike that a similar connection formula holds on Stokes curves 
emanating from simple poles (\cite{Koike2000}). 
It should be noted that, when describing this formula in a general form, 
it is possible to add correction terms in $\hbar$ to the potential function $Q$ as follows. 

\smallskip
\begin{ass} ~ 
\begin{itemize}
\item[({\rm i})] 
The potential function $Q$ in \eqref{eq:sch} takes the form
\begin{equation}
Q = Q_0(x) +\hbar^{2} Q_2(x),
\end{equation} 
where $Q_0$ has a simple pole at $v$, while $Q_2$ has 
at most double pole there. 

\item[({\rm ii})] 
Let $C$ be a Stokes curve emanating from $v$ 
which forms a common boundary of two adjacent Stokes regions ${\rm I}$ and ${\rm II}$.
From $v$, region ${\rm II}$ appears next to region ${\rm I}$ in the counter-clockwise direction.

\item[({\rm iii})]
The Stokes graph of \eqref{eq:sch} does not contain any saddle connection.

\end{itemize}
\end{ass}

As well as the Voros connection formula, we take the WKB solution $\psi_{\pm}$ normalized at $v$, 
which is defined by exactly the same manner as \eqref{eq:WKB-TP}.
Then we have the following.


\begin{thm}[{\cite{Koike2000}}] \label{thm:koike-conn}
Under the above setting, the analytic continuation of 
$\Psi_{\pm}^{\rm I}$ to the Stokes region ${\rm II}$ across the Stokes curve $C$ 
is described as
\[
(\Psi_{+}^{\rm I}, \Psi_{-}^{\rm I}) 
= 
(\Psi_{+}^{\rm II}, \Psi_{-}^{\rm II}) \cdot S, 
\]
\begin{equation} \label{eq:Stokes-Koike}
S = \begin{cases}
\begin{pmatrix} 1 & 0 \\ 2{\rm i} \cos(\pi \sqrt{1 + 4\lambda}) & 1 \end{pmatrix} & \quad
\text{if $\displaystyle \int^x_{v} \sqrt{Q_0(x)} \, {\rm d}x > 0$ on $C$}, 
\\[+1.5em]
\begin{pmatrix} 1 & 2{\rm i} \cos(\pi \sqrt{1 + 4\lambda}) \\ 0 & 1 \end{pmatrix} & \quad
\text{if $\displaystyle \int^x_{v} \sqrt{Q_0(x)} \, {\rm d}x < 0$ on $C$},
\end{cases}
\end{equation}
where 
\begin{equation}
\lambda = \lim_{x \to v} (x-v)^2 Q_2(x).
\end{equation}
\end{thm}

From the perspective of exact WKB analysis, a simple pole is interpreted as a turning point, 
whereas it is a regular singular point of \eqref{eq:sch} from the perspective of differential equations. 
Therefore, the connection multiplier in \eqref{eq:Stokes-Koike} includes a quantity 
related to the monodromy matrix around the regular singular point (i.e., the characteristic exponent).

\bigskip
\begin{exc} \label{exc:simple-pole-normal-form}
Prove the connection formula \eqref{eq:Stokes-Koike} 
for the Schr\"odinger-type ODE \eqref{eq:sch} with the potential
\begin{equation}  \label{eq:Bessel-potential}
Q = \frac{1}{x} + \hbar^2 \frac{\lambda}{x^2}
\end{equation}
by following the steps below:
\begin{itemize}
\item[(1)]
Show that the Borel transform of the WKB solution normalized at the simple pole $x=0$ are given as 
\begin{equation} \label{eq:bessel-borel-2f1}
\begin{cases}
\displaystyle 
\psi_{+, B}(x,\zeta) = C \, w^{- \frac{1}{2}} 
{}_{2}F_{1}\Bigl( \alpha - \frac{1}{2}, \beta - \frac{1}{2}, \frac{1}{2}; w \Bigr), 
\\[+1.0em]
\displaystyle
\psi_{-, B}(x,\zeta) = C \, (w-1)^{- \frac{1}{2}} 
{}_{2}F_{1}\Bigl( \alpha - \frac{1}{2}, \beta - \frac{1}{2}, \frac{1}{2}; 1 - w \Bigr).
\end{cases}
\end{equation} 
with 
\begin{equation}
C = \frac{1}{2\sqrt{\pi}}, \quad 
w = \frac{\zeta}{4 \sqrt{x}} + \frac{1}{2},
\end{equation} 
and $\alpha, \beta$ are constants satisfying $\alpha + \beta = 2$ and $\alpha \beta = - 4\lambda$.

\item[(2)] Following the discussion in \S \ref{subsec:voros-formula}, 
find all singularities of $\psi_{\pm, B}(x,\zeta)$ on the Borel plane, 
as well as the discontinuity at that point, and derive the connection formula \eqref{eq:Stokes-Koike}.

\end{itemize}
\end{exc}

\subsection{Description of monodromy/Stokes matrices via Voros periods}

The connection formulas discussed in \S \ref{sec:conn-formula} are highly effective 
for analyzing the global properties of solutions to differential equations, 
specifically for addressing the {\em direct monodromy problem}.
Here, we will briefly explain the content of \cite[\S 3]{KT98}; 
that is, the calculation of the monodromy matrix/Stokes matrix of 
the Schr\"odinger-type ODE \eqref{eq:sch} using the period integrals, 
which we call the Voros periods, on the spectral curve $\Sigma$.

Here, we will introduce the result of calculating the Stokes matrix around 
the irregular singular point $x = \infty$ for the Weber equation \eqref{eq:Weber-eq}. 
This example is quite simple, and the Stokes matrix can be calculated 
without using the theory of exact WKB analysis. 
However, from the perspective of understanding how the period integrals arise, 
it is also an essential example. The monograph \cite[\S 3]{KT98} addresses 
more nontrivial examples that truly require the theory of exact WKB analysis, 
so we recommend referring to it as well.

\begin{figure}[t]
  \begin{center} 
  \includegraphics[width=60mm]{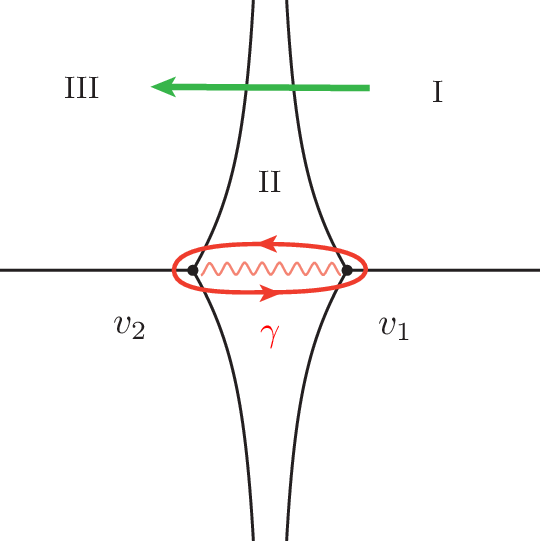} 
  \end{center}  
  \caption{The Stokes graph of the Weber equation for $\nu > 0$ (black) 
  and the closed cycle $\gamma$ (red).}
  \label{fig:weber-Stokes}
 \end{figure}

For simplicity, we consider the case $\nu \in {\mathbb R}_{>0}$.  
In this case, the Stokes graph of the Weber equation \eqref{eq:Weber-eq} 
is given in Figure \ref{fig:weber-Stokes}. 
$x = \infty$ is an irregular singular point of Poincar\'e rank 2, with four singular directions. 
These correspond to the directions in which the Stokes curves asymptotically approach. 
For example, finding the Stokes matrix for the singular direction $\arg x = \pi/2$ 
is equivalent to determining the connection matrix between the Stokes regions ${\rm I}$ and ${\rm III}$. 
Let's carry this out.

Here we take the WKB solutions normalized at $x = v_1 = + 2 \sqrt{\nu}$:
\begin{equation} \label{eq:wkb-v1}
\psi_{\pm}(x,\hbar) = \frac{1}{\sqrt{P_{\rm odd}(x,\hbar)}}
\exp\left( \pm \int^{x}_{v_1} P_{\rm odd}(x,\hbar) \, {\rm d}x \right).
\end{equation}
In computing the analytic continuation from the region ${\rm I}$ to ${\rm III}$
passing through ${\rm II}$, we use the Voros formula twice. 
At the first crossing, we can simply use Theorem \ref{thm:voros-conn}:  
\begin{equation} \label{eq:weber-conne-1}
(\Psi^{\rm I}_+, \Psi^{\rm I}_-) = (\Psi^{\rm II}_+, \Psi^{\rm II}_-) \cdot 
\begin{pmatrix} 1 & {\rm i} \\ 0 & 1 \end{pmatrix}. 
\end{equation}
In discussing the connection problem at the second crossing, 
a care must be taken with the normalization of the WKB solution. 
In order to apply Theorem \ref{thm:voros-conn} at the second crossing point, 
it is necessary to take the WKB solution 
\begin{equation}
\widetilde{\psi}_{\pm}(x,\hbar) = 
\frac{1}{\sqrt{P_{\rm odd}(x,\hbar)}}
\exp\left( \pm \int^{x}_{v_2} P_{\rm odd}(x,\hbar) \, {\rm d}x \right)
\end{equation}
normalized at the turning point $v_2$ where the Stokes curve, 
which we are attempting to cross, emanates.
When replacing the lower endpoints of the integration in the WKB solution, 
the exponential of the period integral emerges as an overall factor as follows:
\begin{equation}
\psi_{\pm}(x, \hbar) = 
\exp\left( \pm \frac{1}{2} V_{\gamma}(\hbar) \right)
\widetilde{\psi}_{\pm}(x,\hbar), 
\qquad
V_\gamma(\hbar) = \oint_{\gamma} P_{\rm odd}(x,\hbar) \, {\rm d}x. 
\end{equation}
Here, $\gamma$ is the closed cycle (which represents a class in $H_1(\Sigma'; {\mathbb Z})$)
which encircles turning points $v_1$ and $v_2$, depicted in red in Figure \ref{fig:weber-Stokes}.  
We note that $V_\gamma$ is Borel summable if the Stokes graph 
does not contain any saddle connection (see Footnote \ref{footnote:saddle}).
Since Theorem \ref{thm:voros-conn} is valid for the WKB solution $\widetilde{\psi}_{\pm}$
on the second crossing point, we have 
\begin{equation} \label{eq:weber-conne-2}
(\Psi^{\rm II}_+, \Psi^{\rm II}_-) = (\Psi^{\rm III}_+, \Psi^{\rm III}_-) \cdot 
\begin{pmatrix} 1 & {\rm i} \, {\rm e}^{-{\mathcal V}_{\gamma}} \\ 0 & 1 \end{pmatrix},
\end{equation}
where ${\mathcal V}_{\gamma}$ is the Borel sum\footnote{\label{foot:weber-period}
In the Weber example, $V_{\gamma}$ is simply given by $- 2 \pi {\rm i} \nu/\hbar$, 
so we do not have take the Borel sum.} 
of $V_{\gamma}$. 
From \eqref{eq:weber-conne-1} and \eqref{eq:weber-conne-2}, 
we have the following resulting connection formula between ${\rm I}$ to ${\rm III}$:
\begin{equation} \label{eq:stokes-coefficient-weber}
(\Psi^{\rm I}_+, \Psi^{\rm I}_-) = (\Psi^{\rm III}_+, \Psi^{\rm III}_-) \cdot 
\begin{pmatrix} 1 & {\rm i} (1 + {\rm e}^{-{\mathcal V}_{\gamma}}) \\ 0 & 1 \end{pmatrix}. 
\end{equation}
Thus we have obtained an explicit expression of the connection matrix for the WKB solution. 

\begin{rem} \label{rem:Stokes-Weber-irreg}
The resulting Stokes matrix \eqref{eq:stokes-coefficient-weber} of the Weber equation 
is different from the one in literatures (e.g., \cite[\S 1]{KT98}). 
Usually, in computing the Stokes matrix around an irregular singular point, 
a formal solution expanded at the the singular point, 
which is different from \eqref{eq:wkb-v1}, is typically used. 
To compare \eqref{eq:stokes-coefficient-weber} with those results, 
we should take the WKB solution normalized at $x=\infty$ 
defined as follows\footnote{
$P_{-1}$ has a singularity at infinity, so the endpoints of the integral were chosen differently. 
By considering a suitably regularized integral, it is also possible to define 
an integral of $P_{-1}$ from infinity; see \cite[Remark 4.6]{IKoT-1}. 
}:
\begin{equation}
\psi_{\pm, \infty}(x,\hbar) = \frac{1}{\sqrt{P_{\rm odd}(x,\hbar)}} \,
\exp\left( 
\pm \hbar^{-1} \int^{x}_{v_1} P_{-1}(x') \, {\rm d}x' 
\pm \int^{x}_{\infty} \bigl( P_{\rm odd}(x',\hbar) - \hbar^{-1} P_{-1}(x') \bigr)\,{\rm d}x' 
\right).
\end{equation}
The difference from \eqref{eq:wkb-v1} is given by 
an alternative integral of $P_{\rm odd}$ which was considered already in 
Exercise \ref{eq:Weber-Bernoulli}: 
\begin{equation} \label{eq:relation-web-wkb-sols}
\psi_{\pm}(x,\hbar) = \exp\left(\pm \frac{1}{2}W(\hbar)\right) \psi_{\pm, \infty}(x,\hbar),
\end{equation}
where 
\begin{equation} \label{eq:web-voros-path}
W(\hbar) = 
\sum_{k \ge 1} \frac{(2^{1-2k}-1) B_{2k}}{2k(2k-1) \, \nu^{2k-1}} \, \hbar^{2k-1}
\end{equation}
is the generating series of the integrals given in \eqref{eq:weber-voros-terms} 
(c.f., \cite[Theorem 1.1]{Takei08}).
The Borel transform of $W(\hbar)$ is explicitly given by
\begin{equation} \label{eq:boreltr-web-voros}
{\mathcal B}W(\zeta) = \frac{1}{2 \zeta} \left(
\frac{1}{\exp(\frac{\zeta}{2\nu}) - 1} + \frac{1}{\exp(\frac{\zeta}{2\nu}) + 1}
- \frac{2 \nu}{\zeta}
 \right),
\end{equation}
where we have used \eqref{eq:def-bernoulli}. 
Then, the Binet's formula implies that the Borel sum of ${\rm e}^{W}$ is  
\[
{\mathcal S} {\rm e}^{W(\hbar)} = 
\frac{{\rm e}^{\nu/\hbar} \, \Gamma(\frac{\nu}{\hbar} + \frac{1}{2})}
{\sqrt{2 \pi} \, (\nu/\hbar)^{\nu/\hbar}}.
\]
Consequently, the $\Gamma$-function appears 
in the Stokes matrices for $\psi_{\pm, \infty}$. 
The resulting formula should be comparable the ones in literatures.

While this may diverge from the main discussion, 
the following remark is significant and thus merits inclusion here 
(for a more detailed discussion, see \cite[\S 2]{Takei08}).
We may observe from \eqref{eq:boreltr-web-voros} that 
${\mathcal B}W(\zeta)$ has the singularities at $\zeta = 2 \pi {\rm i} \nu k$
with $k \in {\mathbb Z}_{\ne 0}$.
Therefore, when $\nu$ is varied into the complex plane, 
$W(\hbar)$ becomes non-Borel summable if the condition 
\begin{equation} \label{eq:non-borel-summable-condition}
2 \pi {\rm i} \nu \in {\mathbb R}_{\ne 0}
\end{equation}
is satisfied.
Actually, the condition \eqref{eq:non-borel-summable-condition} 
is precisely the condition for saddle connections to appear in the Stokes graph of the Weber equation. 
This can be understood from the fact that the values of the period integrals of 
$\sqrt{Q(x)} \, {\rm d}x$, which define the Stokes curves, along $\gamma$ is given by 
$-2 \pi {\rm i} \nu$.
In view of \eqref{eq:relation-web-wkb-sols}, 
the WKB solution $\psi_{\pm}$ is not Borel summable as well if 
the condition \eqref{eq:non-borel-summable-condition} is satisfied. 
When the integration path cannot be separated from the saddle connection using Koike's trick, 
there are cases where Borel summability does not hold. 
Therefore, the assumption regarding the absence of saddle connections in 
Theorem \ref{thm:summablity} is essential. 
\end{rem}

As we have seen in \eqref{eq:stokes-coefficient-weber}, the period integral of $P_{\rm odd}$ 
naturally appears in the expression 
as a consequence of multiple use of the Voros connection formula.  
This is also true for more general examples. 
Those periods are crucially important objects 
not only in the exact WKB analysis, but also in the theory of 
ordinary differential equations on the complex domain. 

\begin{dfn}
The formal series defined by the term-wise period integral\footnote{
There is another type of the Voros periods, 
namely those associated with a relative homology classes 
in $H_1(\overline{\Sigma}, \overline{\Sigma}\setminus \Sigma; {\mathbb Z})$. 
The path in Figure \ref{fig:weber-path} represents such a class for the Weber example. 
These two types of Voros periods are both important in applications; 
see \cite{IN14, IN15} for an application to cluster algebras.  
\label{foot:voros-periods-relative}
} 
\begin{equation}
V_{\gamma}(\hbar) = \oint_{\gamma} P_{\rm odd}(x, \hbar) \, {\rm d}x
\quad (\gamma \in H_1(\Sigma', {\mathbb Z}))
\end{equation}
is called the {\em Voros period} for the cycle $\gamma$. 
\end{dfn}

In summary, we have 

\begin{thm}[{\cite[\S 3]{KT98}; see also \cite{SAKT91}}] \label{thm:voros-period}
Suppose that the Stokes graph of the Schr\"odinger-type ODE \eqref{eq:sch} 
doesn't contain any saddle connection. 
Then, the monodromy matrices and Stokes matrices of \eqref{eq:sch} 
(with respect to the basis of solutions given by the Borel sum of WKB solutions)
are explicitly described by the Borel sum of the Voros periods. 
\end{thm}

The spectral curve of the Weber equation \eqref{eq:Weber-eq} discussed above is of genus zero. 
As a result, the Voros period expressed 
in the aforementioned formula \eqref{eq:stokes-coefficient-weber} 
can actually be written explicitly as the residue at $x = \infty$ 
(c.f., Footnote \ref{foot:weber-period}). 
The same thing happens with {\em Gauss's hypergeometric differential equation}. 
The Schr\"odinger-type ODE \eqref{eq:sch} 
with the $\hbar$-depending potential 
\begin{equation} \label{eq:gauss-HG}
Q = 
\left( \frac{\theta_0^2}{x^2} + \frac{\theta_1^2}{(x-1)^2} 
+ \frac{\theta_\infty^2 - \theta_0^2 - \theta_1^2}{x (x-1)} \right)
+ \hbar^2 \left(
- \frac{1}{4x^2} - \frac{1}{4(x-1)^2} + \frac{1}{4x(x-1)} \right)  
\end{equation}
is an equivalent equation.
For this example, any Voros period along a closed path 
can be expressed using the residues of 
$P_{\rm odd}(x,\hbar)\,{\rm d}x$ at the regular singular points $x=0$, $1$ and $\infty$, 
which is identical to $\theta_0/\hbar$, $\theta_1/\hbar$ and $\theta_\infty/\hbar$, respectively (up to sign).
As we will see in Exercise \ref{exc:characteristic-exp} below, 
these residues are essentially given by the characteristic exponents. 
In other words, the important property of Gauss's hypergeometric differential equation, 
where ``the monodromy matrices can be written using only the characteristic exponents", 
can be geometrically understood as a consequence of the fact that 
``the genus of the spectral curve of \eqref{eq:gauss-HG} is 0"\footnote{
This property also make the computation of Voros period for relative cycles easier,
and one can obtain an explicit expression in terms of 
the Bernoulli numbers such as \eqref{eq:web-voros-path}; 
see \cite{AT13, ATT21, IKoT-1, IKoT-2} for details.
}.

However, in general, when the spectral curve has genus greater than $0$, 
the Voros period becomes a generating series of (hyper-)elliptic integrals, 
and the entries of the monodromy matrix, described as the Borel sum of them, 
become highly transcendental objects. 
We consider this ``description of the monodromy data through period integrals (Voros periods)" 
as an important guiding principle in the exact WKB analysis, 
and in the next Part II, 
we will explore its application to the Painlev\'e equations through the isomonodromy deformation.

\begin{exc} \label{exc:characteristic-exp}
Let us consider the Schr\"odinger-type ODE \eqref{eq:sch} 
with an $\hbar$-depending potential 
\begin{equation}
Q = Q_0(x) + \hbar Q_1(x) + \hbar^2 Q_2(x)
\end{equation}
and assume that each $Q_i(x)$ has order 2 pole at $x=0$
(i.e., $x=0$ is a regular singular point). 

\begin{itemize}
\item[(1)]
For each $m \ge -1$, prove that $P^{(\pm)}_m(x)$ has a pole at $x=0$ 
of order at most $1$.

\item[(2)]
Prove that the residues 
\begin{equation} 
\rho^{(\pm)}(\hbar) = \Res_{x=0} P^{(\pm)}(x,\hbar) {\rm d}x 
~~\left( = \sum_{m \ge -1} \hbar^m \Res_{x=0} P^{(\pm)}_m(x) {\rm d}x \right)
\end{equation} 
are convergent series of $\hbar$ and gives the characteristic exponents 
of \eqref{eq:sch} at $x=0$ 
(namely, the eigenvalues of local monodoromy matrix around $x=0$ is given by 
$\exp(2 \pi {\rm i} \rho^{(\pm)}(\hbar))$). 
\end{itemize}

\end{exc}

\smallskip
\subsection{Topics related to Part I}
\label{subsec:comments-wkb}


\begin{itemize}
\item 
As with the original WKB method, the exact WKB method also has numerous applications in quantum mechanics; 
see \cite{Voros83, DDP93, DDP97-a, DDP97-b} for example.
It is also worth emphasizing that the exact WKB method can be applied 
to quantum mechanical problems that are difficult to analyze by more standard methods.
In particular, in relatively recent work, interesting connections have been discovered 
with topological string theory (in the so-called Nekrasov--Shatashvili background) 
and the refined holomorphic anomaly equations, 
providing new methods for the non-perturbative analysis of quantum mechanics.
For research in this direction, see \cite{CM16, CMS17, vSV23, gm22-b}, 
as well as the Mari\~no's Les Houches Lecture Note \cite{Marino-les-houches-24}.

\item 
The note has dealt only with Schr\"odinger-type ODEs, 
but part of the theory of the exact WKB analysis has also been extended 
to {higher-order ODEs} of the following form: 
\begin{equation} \label{eq:higher-ODE}
\left( 
\hbar^{n} \frac{{\rm d}^n}{{\rm d}x^n} + q_1(x) \hbar^{n-1} \frac{{\rm d}^{n-1}}{{\rm d}x^{n-1}} 
+ \cdots + q_{n-1}(x) \hbar  \frac{{\rm d}}{{\rm d}x} +  q_n(x)
\right) \psi(x,\hbar) = 0.
\end{equation}
Unlike the second-order case, it has been shown by Berk--Nevins--Roberts (\cite{BNR}), 
that discontinuities in the WKB solutions can occur even on 
{\em new Stokes curves} that do not originate from usual turning points. 
Building on their considerations, Aoki--Kawai--Takei proposed 
a candidate for the Stokes graph of higher-order ODEs \cite{AKT94, AKT01}
(see also \cite{Honda-Kawai-Takei}). 
However, a general theorem on the Borel summability of WKB solutions 
has not yet been established\footnote{Recently, a related result is proposed in \cite{Nemes-23}.}. 
For the direction, Aoki--Kawai--Takei \cite{AKT01} and Koike--Takei \cite{KT13} 
have succeeded in providing a novel geometric interpretation of 
the new Stokes curves based on the {\em exact steepest descent method}. 
Generalizing their ideas and giving a rigorous proof 
of Borel summability and connection formulas, 
as well as extending the results shown in this note to higher-order ODEs, 
remain significant and challenging research problems.

\item 
The (candidate of) Stokes graphs of higher-order ODEs were rediscovered 
in the study of {\em BPS states} in four-dimensional ${\mathcal N}=2$ supersymmetric field theories, 
called the theories of class S, by Gaiotto--Moore--Neiztke in \cite{GMN12}, 
and are also known as {\em spectral networks}. 
They also proposed the notion of {(non-)abelianization}; 
a relation between an $SL_n({\mathbb C})$-connection on the base Riemann surface 
and a $GL_1({\mathbb C})$-connection on the $n : 1$ covering (\cite{GMN12, HN13}). 
When $n=2$, this can be understood as the geometric formulation of 
the result of Theorem \ref{thm:voros-period}. 
See \cite{HK17, HN13,  HN19, KKRTT24} for more applications.

\item 
The Schr\"odinger-type ODE with the potential \eqref{eq:Bessel-potential}
with $\lambda = - 1/4$ is equivalent to the {\em quantum differential equation}
for ${\mathbb P}^1$; that is, the differential equation satisfied by 
the {Givental $J$-function} for ${\mathbb P}^1$ (\cite[\S 3]{Giv95}). 
The standard quantum (K\"ahler) parameter $q = {\rm e}^{t}$ is identified with $x$ in this note. 
For this example, the Stokes multiplier $2{\rm i} \cos(\pi \sqrt{1 + 4\lambda}) = 2{\rm i}$, 
computed by the Koike's connection formula, 
is identical to the Euler pairing $\chi({\mathcal O}, {\mathcal O}(1))$
in the derived category $D^{b} {\rm Coh} ({\mathbb P}^1)$ 
of coherent sheaves on ${\mathbb P}^1$ (up to the trivial factor ${\rm i}$). 
This is consistent with the celebrated {\em Dubrovin's conjecture} \cite{dubrovin98, Guzzetti}.
See also \cite[\S 6]{FIMS} for a related discussion for 
the Stokes matrices of the {equivariant} ${\mathbb P}^1$. 

\item 
The explicit formula \eqref{eq:boreltr-web-voros} of the Borel transform of $W(\hbar)$, 
discussed in Remark \ref{rem:Stokes-Weber-irreg}, 
allows us to describe an explicit formula of the Stokes phenomenon for $W(\hbar)$. 
If we denote by ${\mathfrak S}$ the Stokes automorphism (c.f., \cite{Ecalle, Sauzin}), 
then its action to  ${\rm e}^{W}$ 
is given by 
\begin{equation} \label{eq:ddp-web}
{\mathfrak S} {\rm e}^{W} = 
\begin{cases}
\displaystyle 
{\rm e}^{W} (1 + {\rm e}^{2 \pi {\rm i} \nu/\hbar}) & \text{if $\nu \in {\rm i}\,{\mathbb R}_{> 0}$}, 
\\[+.5em]
\displaystyle 
{\rm e}^{W} (1 + {\rm e}^{- 2 \pi {\rm i} \nu/\hbar})^{-1} & \text{if $\nu \in {\rm i} \,{\mathbb R}_{< 0}$}.
\end{cases}
\end{equation}
See \cite[\S 2]{Takei08} for the derivation of the formula based on an explicit alien calculus.  
In fact, similar discontinuities generally happens, as shown in \cite{Voros83, DDP93, IN14}.
The general formula describing the Stokes phenomenon for the Voros period is known as 
the {\em Delabaere--Dillinger--Pham (DDP) formula}, 
and \eqref{eq:ddp-web} serves as an example of it.
In fact, the DDP formula has a close relationship with 
the {(generalized) cluster algebras}, {higher Teichm\"uller theory} and {BPS structures}. 
See \cite{AHT22, All18, AB18, Bri16, BS, GMN09, GMN12, IN14, IN15,  KW24} for more details. 
DDP formula will play a crucial role in deriving a (conjectural) 
connection formula for Painlev\'e $\tau$-function in Part II.

\item 
As we have mentioned in Example \ref{ex:airy}, through the TR/QC correspondence, 
the coefficients of the WKB solutions of Sch\"odinger-type ODEs 
are related to the quantities computed by the {\em topological recursion (TR)} of \cite{EO07} in many examples.
For the Airy equation \eqref{eq:airy-eq}, 
the relationship is described as follows:
\begin{align}
& \psi_{\pm}^{\rm Airy}(x,\hbar) \notag \\
& ~ = \exp\left( \sum_{\substack{g \ge 0 \\  n \ge 1}} \frac{(\pm \hbar)^{2g-2+n}}{n!} 
\int^{z(x)}_{\infty} \hspace{-1.em} \cdots \int^{z(x)}_{\infty} \biggl( W^{\rm Airy}_{g,n}(z_1, \dots, z_n) - 
\delta_{g,0}\delta_{n,2} \frac{{\rm d}x(z_1) \,{\rm d}x(z_2)}{(x(z_1) - x(z_2))^2} \biggr) \right),
\label{eq:airy-tr}
\end{align}
where $W_{g,n}^{\rm Airy}$ is the topological recursion correlator defined from the initial data
\begin{equation} \label{eq:tr-spcurve-airy}
(C = {\mathbb P}^1, x(z) = z^2, y(z) = z), 
\end{equation}
which gives a parametrization of the the spectral curve $y^2 = x$ of the Airy equation. 
We will briefly recall the definition of correlators in \S \ref{subsec:tr-def}. 
It is known that the $W_{g,n}^{\rm Airy}$ is related to the so-called 
$\psi$-class intersection numbers as 
\begin{equation}
W^{\rm Airy}_{g,n}(z_1,\dots,z_n) = \frac{1}{2^{2g-2+n}} \sum_{d_1, \dots, d_n \ge 0} 
\left(\int_{\overline{\mathcal M}_{g,n}} \psi_1^{d_1} \cdots \psi_n^{d_n} \right)
\prod_{i=1}^{n} \frac{(2d_i+1)!!}{z_i^{2d_i+2}} \, {\rm d}z_{i}
\label{eq:airy-wgn-moduli} 
\end{equation}
for $2g-2+n \ge 1$. 
See \cite{Eyanrd11}, and the Les Houches Lecture Note \cite{Gia-Lew} by Giacchetto--Lewanski 
for the definition of the $\psi$-class intersection numbers.
The correspondence offer several new applications to enumerative problems 
which can be handled with TR. Let us recall some of them. 

\smallskip
For the Airy equation, we can completely describe 
the singularity structures on the Borel plane by the formula \eqref{eq:airy-borel-2f1}.
Then, the Darboux's method implies that the location of the Borel singularities 
and the Stokes multipliers are related to the growth order of the coefficients of the WKB solutions.  
In this Airy case, 
the coefficients are related to the intersection numbers on ${\mathcal M}_{g,n}$
as shown in \eqref{eq:airy-tr}--\eqref{eq:airy-wgn-moduli},
and consequently, 
one can describe the asymptotic expansion of the intersection numbers 
on ${\mathcal M}_{g,n}$ when $g \to \infty$. 
See \cite{EGFGGL} for such a new application of TR/QC correspondence. 

On the other hand, the series $W(\hbar)$ in \eqref{eq:web-voros-path} 
is one of the Voros period of the Weber equation \eqref{eq:Weber-eq} 
for a relative cycle on the spectral curve (c.f., Footnote \ref{foot:voros-periods-relative}). 
In \cite{IKoT-1, IKoT-2}, for a class of examples, 
it was proved that the Voros period is related to 
the free energy of the topological recursion through the TR/QC correspondence. 
From this perspective, one can derive a well-known formula 
\begin{equation} \label{eq:HZ-P-N}
F^{\rm Weber}_g = \frac{B_{2g}}{2g(2g-2) \, \nu^{2g-2}} \quad (g \ge 2) 
\end{equation}
as a consequence of the formula \eqref{eq:web-voros-path} for the Voros period $W(\hbar)$.
Here, $F^{\rm Weber}_g $ is the $g$-th free energy of 
\begin{equation} \label{eq:tr-spcurve-weber}
\left( C = {\mathbb P}^1, x(z) = \sqrt{\nu}(z+z^{-1}), y(z) = \frac{\sqrt{\nu}}{2}(z-z^{-1}) \right), 
\end{equation}
which is a parametrization of the spectral curve of the Weber equation \eqref{eq:Weber-eq}.
Originally, the formula \eqref{eq:HZ-P-N} was proved in \cite{HZ, Norbury09, Penner88} 
through its geometric interpretation; 
that is, $F^{\rm Weber}_g$ is related to the Euler characteristic of 
the moduli space ${\mathcal M}_g$ of Riemann surfaces.
From the special function-theoretic perspective, 
these Bernoulli numbers in \eqref{eq:web-voros-path}
can be interpreted as arising from the asymptotic expansion of the $\Gamma$-function 
(i.e., the Stirling expansion) of the Stokes coefficient at the irregular singular point
of the Weber equation (see Remark \ref{rem:Stokes-Weber-irreg}).
In this way, the TR/QC correspondence can be effective in providing 
a special function-theoretic interpretation 
for enumerative invariants and in deriving explicit formulas for them. 
This is also one of the important applications of TR/QC correspondence. 

TR will also be essentially used to study Painlev\'e equations in Part II, 
and we will also provide further (conjectural) applications of TR/QC correspondence.

\end{itemize}

\newpage
\section{Part II : Application to the first Painlev\'e Equation}

In Part II, we discuss the (potential) applications of the theory of 
exact WKB analysis to the Painlev\'e equations. 
We consider the simplest one only, 
called the {\em first Painlev\'e equation}\footnote{
The original Painlev\'e equations do not contain $\hbar$ 
and can be obtained by simply setting $\hbar = 1$. 
We put the parameter $\hbar$ to apply the exact WKB analysis. 
It can be added through a rescaling of variables, and hence, 
analysis for small $\hbar$ is related to analysis for large $t$ 
in the original variable. 
There are many results on large $t$ asymptotics of Painlev\'e I; 
see \cite{FIKN} for example.   
}:
\begin{equation}
(P_{\rm I}) ~:~
\hbar^2 \frac{{\rm d}^2q}{{\rm d}t^2} = 6q^2 + t.
\end{equation}
The fact that the exact WKB analysis is effective in describing monodromy 
naturally leads to exploring its application to Painlev\'e equations through 
the theory of {\em isomonodromy deformations} of linear ODEs\footnote{
As is mentioned in introduction, Aoki, Kawai, and Takei pioneered 
the theory of exact WKB analysis for the Painlev\'e equations. 
Using a different approach from the one introduced here, 
they have already constructed formal series solutions (called ``instanton-type solutions'') 
containing two arbitrary constants. 
The relationship between these two approaches will be discussed \S \ref{subsec:comments-part2}.
}. 
However, as explained in Section \ref{section:naive-spectral-curve}, 
a naive spectral curve of isomonodromic linear ODE is ill-defined. 
To select an appropriate spectral curve, insights from exact WKB analysis are essential. 
For a suitably chosen spectral curve, exact WKB analysis, topological recursion, 
and isomonodromy deformations intertwine harmoniously, 
providing a new method for constructing 
the $\tau$-function associated with the general solutions of Painlev\'e equation 
is constructed as a formal series in $\hbar$. 
While the Borel summability of these formal series has not been rigorously proven, 
assuming its validity leads to highly explicit connection formulas for $\tau$-functions, 
which agrees with several known results. 
These, in turn, yield explicit conjectural formulas concerning the resurgent structure 
of the partition functions for the topological recursion. 
The purpose of Part II is to introduce these cutting-edge research developments in mathematical physics.

The method explained in Part II cannot be applied directly to 
arbitrary nonlinear ordinary differential equations. 
Our approach relies essentially on the integrability of the Painlev\'e equations, 
namely, their characterization in terms of isomonodromic deformations. 
Such a distinctive feature of the Painlev\'e equations allows a wide variety of analytic, 
geometric, and algebraic methods to be applied despite the nonlinear nature of the equations, 
making it possible to study their singularities, asymptotic behavior, symmetries, 
and related properties in a unified manner.

\subsection{Isomonodromy deformation and $\tau$-function}

All Painlev\'e equations describe a compatibility condition 
of a certain system of linear PDEs (c.f., \cite{JMU}). 
For the case of $(P_{\rm I})$, the system is given as follows:
\begin{align} \label{eq:lax}
\begin{cases}
\displaystyle 
(L_{\rm I})  ~:~ \left[ \hbar^2 \frac{\partial^2}{\partial x^2} - 
\frac{\hbar}{x-q} \, \Bigl( \hbar \frac{\partial}{\partial x} - p \Bigr)
- (4x^3 + 2t x + 2 H) \right] \psi = 0, \\[+1.em]
\displaystyle 
(D_{\rm I}) ~:~ \left[ \hbar \frac{\partial}{\partial t} - \frac{1}{2(x-q)} \,
\Bigl( \hbar \frac{\partial}{\partial x} - p \Bigr)
\right] \psi = 0,
\end{cases}
\end{align}
where
\begin{equation} \label{eq:Hamiltonian-PI}
H = \frac{p^2}{2} - 2q^3 - t q.
\end{equation}
The compatibility condition of the system \eqref{eq:lax} 
is given by the 
Hamiltonian system
\begin{equation} \label{eq:Ham-PI}
\hbar \frac{{\rm d}q}{{\rm d}t} = \frac{\partial H}{\partial p}, \quad 
\hbar\frac{{\rm d}p}{{\rm d}t} = - \frac{\partial H}{\partial q},
\end{equation} 
which is equivalent to $(P_{\rm I})$.

$(L_{\rm I})$ is a linear ODE with an irregular singular point at $x=\infty$. 
The expression \eqref{eq:Hamiltonian-PI} of $H$ gurantees that 
$x = q$ is an apparent singularity of $(L_{\rm I})$; 
that is, all solutions are holomorphic and have no monodoromy around the point. 
Therefore, the Stokes multipliers around $x = \infty$ 
become the essential monodromy data of $(L_{\rm I})$. 
When the system $(L_{\rm I})$--$(D_{\rm I})$ are compatible, 
the Stokes multipliers of the fundamental solutions remain independent of $t$, 
providing the conserved quantities of $(P_{\rm I})$. 
In fact, one can show that only two of the Stokes multipliers are independent 
(see \eqref{eq:cyclic} below), 
and they parameterize the general solution of $(P_{\rm I})$.
Such the existence of enough numbers of conserved quantity can be understood as 
the {integrability} of Painlev\'e equations. 
Through the detailed analysis of these Stokes multipliers, 
numerous non-trivial properties of the solutions to the Painlev\'e equations are revealed.
For example, \cite{Kap89, Kitaev, FIKN} provides a detailed analysis of the behavior of solutions 
to the Painlev\'e equations using 
the Riemann-Hilbert method.

For a given solution $q(t)$ of $(P_{\rm I})$, 
the associated {\em ${\tau}$-function} $\tau(t)$ 
is defined (up to constant) as a function satisfying 
\begin{equation}
\hbar^2 \frac{{\rm d}}{{\rm d}t} \log \tau(t) = H(t),
\end{equation}
where the right hand side is the corresponding 
Hamiltonian function \eqref{eq:Hamiltonian-PI}. 
It is known that $\tau(t)$ is an entire function
even though the solution $q(t)$ is meromorphic. 
The relationship between a solution of $(P_{\rm I})$
and the $\tau$-function is quite analogous to 
that between the Weierstrass $\wp$-function and 
$\sigma$-function ($\vartheta$-function).

\begin{exc} ~
\begin{itemize}
\item[(1)]
Derive the equation \eqref{eq:Ham-PI} 
from the compatibility condition of the system \eqref{eq:lax}.

\item[(2)]
Suppose that $a \in {\mathbb C}$ is a pole a solution $q(t)$ of $(P_{\rm I})$. 
Compute the first several terms of the Laurent series expansion of $q(t)$ at $t = a$, 
and observe that the coefficient of $(t-a)^{4}$ can be chosen as an arbitrary constant. 
Also, observe that 
the function $H(t)$ behaves as 
\begin{equation}
H(t) = \frac{\hbar^2}{t-a} \left(1 + O(t-a) \right)
\end{equation}
when $t \to a$, and the corresponding $\tau$-function
has a simple zero at $a$.
\end{itemize}
\end{exc}

\subsection{An observation on the spectral curve from the exact WKB perspective}
\label{section:naive-spectral-curve}

As we have seen in the previous subsection, the Stokes multipliers of $(L_{\rm I})$ 
become key to analyzing the properties of the general solutions of $(P_{\rm I})$. 
When attempting to describe those based on the ideas of the exact WKB analysis, 
it is expected that the period integrals (Voros periods) on the spectral curve 
\begin{equation} \label{eq:naive-classical-limit}
y^2 = 4x^3 + 2 t x + 2``H|_{\hbar = 0}"
\end{equation}
of $(L_{\rm I})$ will describe those Stokes multipliers. 

However, certain issues arise here. 
Since $H$ is expressed using the solution of $(P_{\rm I})$, 
it is necessary to understand the behavior of the solution of $(P_{\rm I})$ as $\hbar \to 0$ 
before describing the spectral curve. But, there is no guarantee that 
the function $H$ has a finite limit as $\hbar \to 0$, 
making it difficult to start from \eqref{eq:naive-classical-limit}.

Let us shift our perspective here. If some meaning could be attached to 
$``H|_{\hbar = 0}"$ mentioned above, equation \eqref{eq:naive-classical-limit} would define 
a family of elliptic curves with the independent variable $t$ of $(P_{\rm I})$ as the deformation parameter. 
By specifying the $A$-cycle and $B$-cycle on these curves, 
the period integrals of $y \, {\rm d}x$ along these cycles are expected to describe 
the two independent Stokes multipliers as the leading term of the Voros periods. 
Therefore, if the Stokes multiplier of $(L_{\rm I})$ are independent of $t$, 
the elliptic curve should be deformed so that the periods of $y \, {\rm d}x$ remain independent of $t$. 
However, since $t$ already appears in the coefficients of equation \eqref{eq:naive-classical-limit}, 
it is impossible to deform the elliptic curve so that all periods of $y \, {\rm d}x$ are independent of $t$, 
which presents a problem with this approach as well.

Therefore, (hoping that the $B$-cycle can be dealt with later)
let's first consider a deformation family of elliptic curves 
such that only the period integral of $y \, {\rm d}x$ along the $A$-cycle is independent of $t$:
\begin{equation} \label{eq:spectral-curve-PI}
\Sigma_{P_{\rm I}} ~:~ y^2 = 4x^3+2tx+u(t,\nu).
\end{equation}
This is a family of elliptic curves obtained by imposing the condition
\begin{equation} \label{eq:def-of-u}
\nu = \frac{1}{2\pi {\rm i}} \oint_A y \,{\rm d}x 
\end{equation}
with another parameter $\nu$ that is independent of $t$. 
The equality \eqref{eq:def-of-u} defines $u(t,\nu)$ locally as an implicit function.
In fact, this family of elliptic curves is the correct object to consider 
when constructing the general solution of $(P_{\rm I})$, as will be explained in subsequent sections.

\subsection{Construction of Painlev\'e $\tau$-functions by topological recursion}

Based on the idea described in the previous subsection, can we construct 
a differential equation whose spectral curve is the elliptic curve $\Sigma_{P_{\rm I}}$?
If such a construction is possible, then
(assuming we can temporarily ignore the issue of $t$-dependence of the $B$-period of $y\,{\rm d}x$) 
it would be close to $(L_{\rm I})$.
In fact, it will be shown that this objective can be achieved 
by the {topological recursion} and {quantum curves} (i.e., TR/QC correspondence).

\subsubsection{Topological recursion for $\Sigma_{P_{\rm I}}$} \label{subsec:tr-def}

Following \cite{I19}, let us we apply the {\em topological recursion (TR)} 
to our specific example $\Sigma_{P_{\rm I}}$.
For the general formalism of the TR, 
see \cite{EO07, EO08, Eynard-book} and 
the Les Houches Lecture Note \cite{bouchard-les-houces} by Bouchard. 

First, we regard \eqref{eq:spectral-curve-PI} as 
an initial data\footnote{
A tuple consisting of a Riemann surface and a meromorphic function defined on it, 
like \eqref{eq:elliptic-spectral-curve-tr}, is referred to as a spectral curve 
in the context of topological recursion. 
Since we have already used this term, we will simply refer to 
the spectral curve in topological recursion as the ``initial data'' in this note.
} 
of the TR: 
\begin{equation} \label{eq:elliptic-spectral-curve-tr}
C = {\mathbb C} / L, \qquad 
x = \wp(z), \qquad 
y = \frac{{\rm d}\wp}{{\rm d}z}(z).
\end{equation}
Here, $L = {\mathbb Z} \, \omega_A + {\mathbb Z} \, \omega_B$ 
is the lattice of periods of ${\rm d}x/y$ on $\Sigma_{P_{\rm I}}$, 
and $\wp(z)$ is the associated Weierstrass $\wp$-function. 
We have already specified the $A$-cycle and $B$-cycle in the previous subsection, 
and the corresponding Bergman bidifferential is given by 
\begin{equation} 
B(z_1, z_2) = \left( \wp(z_1 - z_2) + \frac{\eta_A}{\omega_A} \right) \,{\rm d}z_1 {\rm d}z_2,
\end{equation}
where 
\begin{equation}
\eta_A = - \oint_A \frac{x {\rm d}x}{y}.
\end{equation}
The ramification points are given by half periods
\begin{equation}
R = \left\{ \frac{\omega_A}{2}, \frac{\omega_B}{2}, \frac{\omega_A+\omega_B}{2} \right\},
\end{equation}
and $\sigma(z) \equiv - z ~ ({\rm mod} \, L)$ 
gives the local involution around the ramification points\footnote{
For the Airy spectral curve \eqref{eq:tr-spcurve-airy}, we have $R = \{0\}$ 
and $\sigma(z) = -z$.
For the Weber spectral curve \eqref{eq:tr-spcurve-weber}, we have $R = \{\pm 1\}$ 
and $\sigma(z) = z^{-1}$. 
For Both examples, the Bergman bidifferential is given by 
$B(z_1, z_2) = {\rm d}z_1 {\rm d}z_2/(z_1-z_2)^2$.
}.

Let $W_{g,n}(z_1, \dots, z_n)$ be the correlators defined by the TR.  
Namely, 
\begin{equation}
W_{0,1}(z) = y(z) \, {\rm d}x(z), \qquad
W_{0,2}(z_1, z_2) = B(z_1, z_2), 
\end{equation}
and for $2g-2+n \ge 1$, we define
\[ 
W_{g,n+1}(z_{0},z_{1},\dots,z_{n}) = 
\sum_{r \in R} \underset{z = r}{\Res} \,
\frac{\int_{w=\sigma({z})}^{w=z} W_{0,2}(z_{0}, w)}{2\bigl( y(z)-y(\sigma({z})) \bigr){\rm d}x(z)} 
\, R_{g,n}(z, z_1,\dots, z_n),
\] 
\vspace{-.5em}
\begin{align}
& R_{g,n}(z, z_1,\dots, z_n) \notag \\ 
& \quad = 
W_{g-1, n+1}(z,\sigma({z}),z_{1},\dots,z_{n}) 
+ \hspace{-.8em} 
\sum'_{\substack{g_{1} + g_{2} = g \cr
I \sqcup J = \{1,\dots,n\} }} W_{g_{1}, 1+|I|}(z,z_{I}) \,
W_{g_{2}, 1+|J|}(\sigma({z}),z_{J}),
\label{eq:top-rec-higher}
\end{align}
where the prime symbol $'$ means that no $W_{0,1}$ appears in the summation.
We also define
\begin{equation} \label{eq:def-symplectic-invariant}
F_g = \frac{1}{2-2g} \, \underset{r \in R}{\sum}
\Res_{z = r} \biggl(\int^{z} W_{0,1}(z) \Biggr) W_{g,1}(z)
\end{equation}
for $g \ge 2$. $F_0$ and $F_1$ are defined but in an alternative way:
\begin{equation}
F_0 = \frac{t u}{5} + \frac{\nu}{2} \oint_B y \, {\rm d}x, \quad
F_1 = - \frac{1}{12} \log (\omega_A^6 {\mathcal D}), 
\end{equation}
where ${\mathcal D} = - 8t^3 - 27 u^2$ is the discriminant.
See \cite{EO07} for properties of $W_{g,n}$ and $F_g$.

\begin{exc}
Check the following identities: 
\begin{equation} \label{eq:matone-sw}
\frac{\partial F_0}{\partial t} = \frac{u}{2}, 
\quad 
\frac{\partial F_0}{\partial \nu} = \oint_B y \, {\rm d}x, 
\quad
\frac{\partial^2 F_0}{\partial \nu^2} = 2 \pi {\rm i} \, \frac{\omega_B}{\omega_A}.
\end{equation}
\end{exc}

\subsubsection{Perturbative quantum curve and formal monodromy}

Let us introduce the following two generating series
\begin{equation} \label{eq:Z-TR}
Z(t,\nu, \hbar) = 
\exp
\Biggl(
\sum_{g \ge 0} \hbar^{2g-2} F_{g}(t, \nu)
\Biggr)
\end{equation}
and 
\begin{align} 
& \chi_{\pm}(x,t,\nu, \hbar) \notag \\
& ~~ = \exp
\Biggl(
\sum_{g \ge 0, n \ge 1} \frac{(\pm \hbar)^{2g-2+n}}{n!} 
\int^{z(x)}_{0} \cdots \int^{z(x)}_{0} 
\left( 
W_{g,n}(z_1,\dots, z_n) - 
\delta_{g,0} \delta_{n,2} \frac{{\rm d}x(z_1) {\rm d}x(z_2)}{(x(z_1) - x(z_2))^2}
\right)
\Biggr), 
\label{eq:chi-pm}
\end{align}
where $z(x)$ is a local inverse function of $x = \wp(z)$.
The first one is called the {\em perturbative partition function}, 
while the second one is called the  {\em perturbative wave function}\footnote{
Both of them are named as ``function", but they are divergent series of $\hbar$ in general. 
The Borel summability of these formal series are not rigorously proved in general. 
Recently, the convergence of the Borel transform is proved in \cite{BEG24}. 
}. 

We can show
\begin{thm}[{\cite[Theorem 3.7 and 3.9]{I19}}] \label{thm:quantum-curve-cubic} ~ 
\begin{itemize}
\item[(i)] ~ 
$\chi_{\pm}$ is a WKB solution of the following PDE: 
\begin{align}
& \biggl[ \hbar^2 \frac{\partial^2}{\partial x^2} 
- 2 \hbar^2 \frac{\partial}{\partial t} - 
\Bigl( 4x^3 + 2t x +  2 \hbar^2 \frac{\partial F}{\partial t}(t,\nu, \hbar) \Bigr) 
\biggr] \chi_{\pm}(x,t,\nu,\hbar) = 0, 
\label{eq:quantum-curve-PDE}
\end{align} 
where $F(t,\nu, \hbar) = \sum_{g \ge 0} \hbar^{2g-2} F_g(t,\nu)$ 
($= \log Z(t,\nu, \hbar)$)
is the total free energy. 

\item[(ii)] ~ 
The term-wise analytic continuation of $\chi_\pm$ along 
$A$-cycle and $B$-cycle are described by 
\begin{equation} \label{eq:monodromy-chi}
\hspace{+3.em}
\chi_{\pm}(x,t,\nu,\hbar) \mapsto 
\begin{cases}
{\rm e}^{\pm 2\pi {\rm i} \nu/\hbar} \, \chi_{\pm}(x,t,\nu,\hbar) & ~~\text{along $A$-cycle}, \\[+1.em]
\dfrac{Z(t,\nu \pm \hbar,\hbar)}{Z(t,\nu,\hbar)}  \,
\chi_{\pm}(x,t,\nu\pm \hbar, \hbar) & ~~\text{along $B$-cycle}. 
\end{cases}
\end{equation}
\end{itemize}
\end{thm}

The equation \eqref{eq:quantum-curve-PDE} is a PDE, and its WKB solutions have not been introduced in this note. 
However, noting that the $\hbar^2$ is multiplied with the partial derivative with respect to $t$, 
it is possible to derive a recursion relation for the coefficients of the WKB-type formal series solution, 
which is similarly to the method described in \S \ref{subsec:wkb-recursion}.
The claim (i) is proved by comparing the recursion relation with the TR.
The claim (ii) is a consequence of the {\em variation formulas} in TR (c.f., \cite[\S 5]{EO07}).

It follows from \eqref{eq:matone-sw} that the classical limit of 
the PDE \eqref{eq:quantum-curve-PDE} is identical to the spectral curve $\Sigma_{P_{\rm I}}$. 
Hence, we call the PDE \eqref{eq:quantum-curve-PDE} the {\em perturbative quantum curve}. 
Also, since $\nu$ is a parameter independent of $t$, we were able to construct 
a differential equation \eqref{eq:quantum-curve-PDE} whose formal monodromy (i.e., term-wise analytic continuation) 
along the $A$-cycle is $t$-independent, as initially intended. 
However, although the perturbative quantum curve \eqref{eq:quantum-curve-PDE} 
obtained in this way is similar in form to the isomonodromic linear ODE $(L_{\rm I})$
(combined with the deformation equation $(D_{\rm I})$), 
it does not completely match. 
We will explain how to fill the gap in the next subsection.

\subsubsection{Construction of the $\tau$-function via discrete Fourier transform}
\label{subsec:kyiv-type-formula}

Although the formal monodromy of $\chi_\pm$ along $A$-cycle is $t$-independent, 
the formal monodromy along the $B$-cycle is described using difference operators. 
In fact, it is known that, in conformal field theory, the analytic continuation 
of a conformal block with a degenerate field insertion can similarly described 
by a shift operator. 
In \cite{ILT}, Iorgov--Lisovyy--Teschner employed the {\em discrete Fourier transform}
to convert the shift operator-valued monodromy into an actual monodromy, 
thereby successfully constructing the $\tau$-function associated with the Painlev\'e VI equation, 
and give an elegant proof of the {\em Kyiv formula} proposed by Gamayun--Iorgov--Lisovyy in \cite{GIL}. 

Let us apply their ideas to our perturbative wave function $\chi_{\pm}$. 
Namely, let us consider the discrete Fourier transform 
of $\chi_\pm$ with respect to the parameter $\nu$ to define
\begin{equation}  \label{eq:NPT-WF}
\psi_{\pm}(x,t,\nu, \rho, \hbar) =  
\frac{\sum_{k \in {\mathbb Z}} {\rm e}^{2\pi {\rm i} k \rho/\hbar} 
Z(t, \nu + k \hbar, \hbar) \,  \chi_{\pm}(x,t,\nu+k \hbar,\hbar)}
{\sum_{k \in {\mathbb Z}} {\rm e}^{2\pi {\rm i} k \rho/\hbar} 
Z(t, \nu + k \hbar, \hbar)},
\end{equation}
which we call the {\em non-perturbative wave function} (c.f., \cite{EM08}). 
Here $\rho$ is a parameter, which is 
dual to $\nu$, assumed to be $t$-independent.  
Actually, the formal series \label{eq:DFT-wave-function} are not a usual power series in $\hbar$, 
but can be regarded as a two-sided trans-series (which contain both positive and negative exponential factors). 
These exponential terms can be summed up to $\vartheta$-functions. 
From the viewpoint of the TR, 
the discrete Fourier transform can be regarded as a non-perturbative correction 
to the perturbative series. See \cite{EM08, I19} for details. 

The previous formal monodromy property \eqref{eq:monodromy-chi} implies
that the term-wise analytic continuation of $\psi_\pm$ becomes 
\begin{equation} \label{eq:formal-t-independent-monodromy}
\hspace{+3.em}
\psi_{\pm}(x,t,\nu,\rho, \hbar)  \mapsto 
\begin{cases}
{\rm e}^{\pm 2\pi {\rm i} \nu/\hbar} \psi_{\pm}(x,t,\nu,\rho, \hbar) & ~~\text{along $A$-cycle}, \\[+.5em]
{\rm e}^{\mp 2\pi {\rm i} \rho/\hbar} \psi_{\pm}(x,t,\nu,\rho, \hbar) & ~~\text{along $B$-cycle}. 
\end{cases}
\end{equation}
In view of the property \eqref{eq:formal-t-independent-monodromy}, 
it is natural to define the 
{\em Voros periods of the non-perturbative wave function} 
along $A$-cycle and $B$-cycle 
by $2 \pi {\rm i} \nu$ and $2 \pi {\rm i} \rho$, respectively, 
even though the parameter $\rho$ is not an actual period integral on the spectral curve. 
Based on the idea of the exact WKB analysis, which states that 
``the Voros periods are fundamental quantities that describe the monodromy/Stokes data'', 
it is natural to expect that $\psi_\pm$ satisfies an isomonodromy system.
In fact, one can prove the following.

\begin{thm}[{\cite[Theorem 4.3 and 4.7]{I19}}] \label{thm:main-theorem} 
The non-perturbative wave function $\psi_\pm$ given in \eqref{eq:NPT-WF}
is a formal solution of the isomonodromy system $(L_{\rm I})$--$(D_{\rm I})$
associated with the first Painlev\'e equation $(P_{\rm I})$, 
where $q$ and $p$ in the system $(L_{\rm I})$--$(D_{\rm I})$ are given by 
\begin{equation} \label{eq:sol-PI}
q(t,\nu,\rho;\hbar) = - \hbar^2 \frac{{\rm d}^2}{{\rm d}t^2} 
\log \tau(t,\nu,\rho, \hbar), 
\quad 
p(t,\nu,\rho;\hbar) = - \hbar^3 \frac{{\rm d}^3}{{\rm d}t^3} 
\log \tau(t,\nu,\rho, \hbar)
\end{equation}
with 
\begin{equation} \label{eq:DFT-tau}
\tau(t,\nu,\rho, \hbar) = 
\sum_{k \in {\mathbb Z}} {\rm e}^{2\pi {\rm i} k \rho/\hbar} 
Z(t, \nu + k \hbar, \hbar).
\end{equation}
Consequently, \eqref{eq:DFT-tau} gives 
a formal series-valued $\tau$-function for $(P_{\rm I})$. 
\end{thm}

The series of the form \eqref{eq:DFT-tau} was first introduced by 
Eynard--Mari\~no in \cite{EM08} as the {\em non-perturbative partition function}. 
Similarly to \eqref{eq:NPT-WF}, 
\eqref{eq:DFT-tau} can be written as a power series in $\hbar$ 
whose coefficients are written by (derivatives of) $\vartheta$-functions. 
Then, we can verify that the first terms of the solution $(q,p)$
of the Hamiltonian system \eqref{eq:Ham-PI} obtained here are given by 
the Weierstrass elliptic functions
associated with the spectral curve $\Sigma_{P_{\rm I}}$:
\begin{equation} \label{eq:boutroux-asymptotic}
\begin{cases}
\displaystyle 
q(t,\nu,\rho,\hbar) = \wp\left( \frac{5t}{4 \hbar} 
+ \Bigl( \frac{\rho}{\hbar} + \frac{1}{2} \Bigr) \omega_A 
+ \Bigl( \frac{\nu}{\hbar} + \frac{1}{2} \Bigr) \omega_B \right) 
+ O(\hbar), \\[+1.5em]
\displaystyle 
p(t,\nu,\rho,\hbar) = \wp'\left( \frac{5t}{4 \hbar} 
+ \Bigl( \frac{\rho}{\hbar} + \frac{1}{2} \Bigr) \omega_A 
+ \Bigl( \frac{\nu}{\hbar} + \frac{1}{2} \Bigr) \omega_B \right) 
+ O(\hbar),
\end{cases}
\end{equation}
where $O(\hbar)$ is not the usual Landau's symbol, 
but it just mean the higher order terms in the $\hbar$-expansion. 
Thus, at the level of formal series computation, 
we have obtained the Boutroux's {\em elliptic asymptotic} found in \cite{Boutroux}. 
It follows from the differential equation satisfied by $\wp$-function that 
\begin{equation}
H(t,\nu,\rho,\hbar) = \frac{u(t,\nu)}{2} + O(\hbar)
\end{equation}
holds. Therefore, we have verified that, the spectral curve $\Sigma_{P_{\rm I}}$ 
precisely agrees with the naive classical limit \eqref{eq:naive-classical-limit} in the above sense.

Theorem \ref{thm:main-theorem} can be simply summarized as 
\[
\text{
``The non-perturbative quantum curve is an isomonodoromy system''.
}
\]  
We also note that it has already been proved in \cite{EG, MO, EGFMO} 
that the $\tau$-functions for other all Painlev\'e equations 
can be constructed from the TR in the same manner 
(see also some related results \cite{IM, IS, BLMST, IMS}).
Thus, while remaining at the level of formal series, 
we have constructed a $\tau$-function for $(P_{\rm I})$ which 
contains two arbitrary parameters $(\nu, \rho)$
by using the TR and discrete Fourier transform. 
This is a TR/exact WKB analogue of the Kyiv formula. 
We will also make several comments on relationship between 
conformal field theory and TR below 
(see  \S \ref{subsec:comments-part2}). 


\bigskip
\subsection{From Painlev\'e  equation to resurgent structure in topological recursion}
\label{subsec:resurgence-painleve}

Next, following \cite[\S 2]{IM23},
let us discuss about the resurgent structure 
of the formal series constructed in the previous section. 
Our strategy is, in studying the resurgent structure of perturbative/non-perturbative partition function,
to use the the Stokes multipliers of $(L_{\rm I})$, 
which is the conserved quantities for the solution of $(P_{\rm I})$. 
In the computation, we will employ the exact WKB analysis which was reviwed in Part I. 
A huge difference from Part I is that, 
since our perturbative quantum curve \eqref{eq:quantum-curve-PDE} is a PDE, 
no rigorous theorem about the Borel summability and connection formulas, 
such as Theorem \ref{thm:summablity}, 
has been established. 
However, in \cite[\S 5]{I19}, by assuming that the results in Part I 
are applicable to our quantum curve \eqref{eq:quantum-curve-PDE}, 
we were able to derive an explicit formula for the resurgent structure. 
Moreover, it turns out that, surprisingly, the formulas obtained by the method perfectly match 
with results obtained through a different approach. 
Let us conclude this note by discussing these observations
(with the hope that those will become rigorous mathematical theorems in the future).

\smallskip
\subsubsection{Conjectures on Borel summability and connection formulas}
\label{subsec:strategy}

\begin{figure}[t]
 \begin{minipage}{0.33\hsize}
  \begin{center}
   \includegraphics[width=45mm]{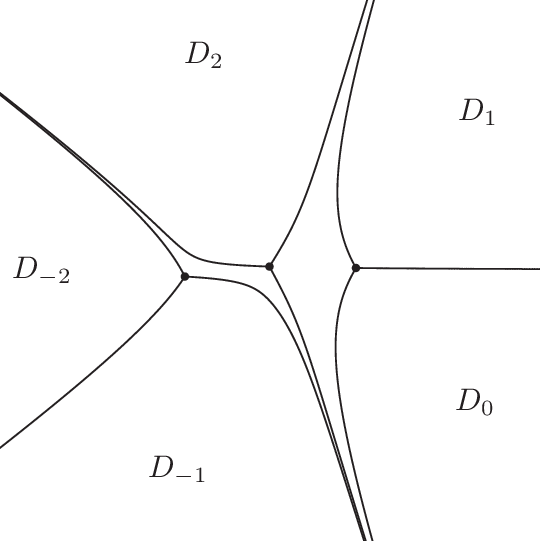} \\
   {$t =  t_c  - {\rm i} \epsilon$}
  \end{center}
 \end{minipage}
  \begin{minipage}{0.33\hsize}
  \begin{center}
   \includegraphics[width= 45mm]{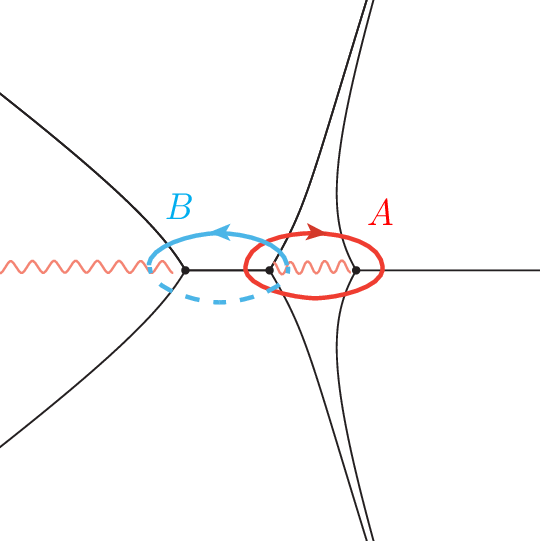} \\
   {$t = t_c$}
  \end{center}
 \end{minipage}
  \begin{minipage}{0.33\hsize}
  \begin{center}
   \includegraphics[width=45mm]{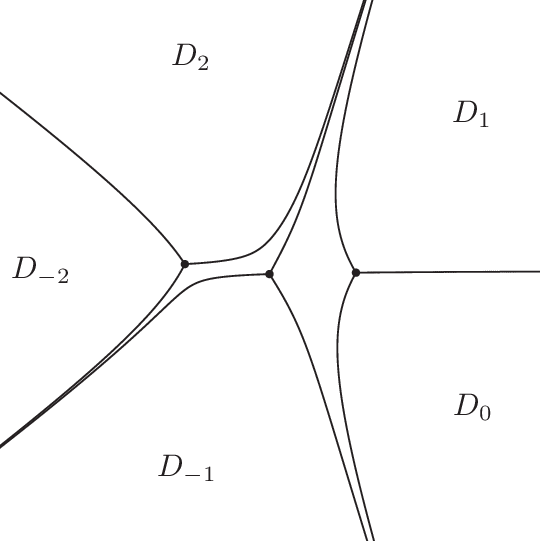} \\
   {$t = t_c + {\rm i} \epsilon$}
  \end{center}
 \end{minipage}
   \caption{Stokes graphs of $\phi$ when $t$ varies near the negative real axis. 
   In this figure, we have chosen $t_c = -5$, $\epsilon=1/2$, and $\nu = 1$. }
  \label{fig:mutation}
\end{figure}

Let us take the meromorphic quadratic differential 
\begin{equation} \label{eq:quad-diff}
\phi(x) = (4x^3 + 2 t x + u(t,\nu)) \, {\rm d}x^2. 
\end{equation}
associated with the spectral curve $\Sigma_{P_{\rm I}}$. 
Figure \ref{fig:mutation} shows Stokes graphs for some values of $t$ and $u(t,\nu)$, 
where we have chosen $A$-cycle and $B$-cycle as is shown in the middle figure. 

The main conjectural ansatz for the following discussion is the following.

\begin{conj}[{\cite[\S 5]{I19}}] ~ \label{conj:hope}
\begin{itemize}
\item[{\rm (i)}] 
If the Stokes graph does not contain any saddle connection, 
then the perturbative partition function \eqref{eq:Z-TR} is Borel summable. 
Moreover, the discrete Fourier series 
\begin{equation} \label{eq:Borel-resumed-tau}
{\mathscr T}(t,\nu,\rho, \hbar) = 
\sum_{k \in {\mathbb Z}} {\rm e}^{2 \pi {\rm i} k \rho/\hbar} {\mathcal Z}(t,\nu+k\hbar, \hbar)
\end{equation}
converges and gives an {\em analytic $\tau$-function} of $(P_{\rm I})$. 
Here, ${\mathcal Z}$ is the Borel sum of the partition function $Z$.  

\item[{\rm (ii)}]
Under the same saddle-free condition, the perturbative wave function $\chi_\pm$  
is Borel summable on each Stokes region. Moreover,  
\begin{equation} \label{eq:Borel-resumed-NPWKB}
{\Psi}_{\pm}(x,t,\nu,\rho, \hbar) 
= \frac{\displaystyle \sum_{k \in {\mathbb Z}} {\rm e}^{2 \pi {\rm i} k \rho / \hbar} 
\, {\mathcal Z}(t, \nu + k \hbar, \hbar) \, {\mathcal X}_{\pm}(x,t,\nu+k \hbar, \hbar)} 
{\displaystyle \sum_{k \in {\mathbb Z}} {\rm e}^{2 \pi {\rm i} k \rho / \hbar}  \, 
{\mathcal Z}(t, \nu + k \hbar, \hbar) } 
\end{equation}
converges and give an analytic solution of the isomonodromy system 
$(L_{\rm I})$--$(D_{\rm I})$ on the Stokes region. 
Here, ${\mathcal X}_{\pm}$ is the Borel sum of $\chi_\pm$. 

\item[{\rm (iii)}]  
Under the same saddle-free condition, the Borel sums 
${\mathcal X}_{\pm}$ defined on adjacent Stokes regions are related by the 
{Voros connection formula} (or the {path-lifting rule} in the sense of \S \ref{subsec:path-lifting}). 
\end{itemize} 
\end{conj}

The claim (i) has already been proposed in \cite{dmp-np, gm22, gkkm23}, 
and the claims (ii), (iii) are inferred from the results from exact WKB analysis for Schr\"odinger-type ODEs
(which was explained in Part I of this note). 
For a class of genus $0$ spectral curves, the conjecture is proved rigorously in \cite{IK-I, IK-II}
where the quantum curve becomes an ODE which can be rigorously handled within 
the general framework introduced in Part I.

Now, since a saddle connection appears on the $B$-cycle when $t$ lies on the negative real axis 
(say $t = - 5$ as in Figure \ref{fig:mutation}), we may expect:
\begin{itemize}
\item 
The Borel transform of the perturbative partition function has 
a singularities on the Borel plane at (integer multiples of) the $B$-periods $\oint_B y \, {\rm d}x$.  
\item 
The singularity 
causes a certain Stokes jump when $t$ crosses the negative real axis. 
\end{itemize}
The main goal of the rest of this note is to derive an explicit formula
which describes the action of the Stokes automorphism. 
The resulting formula will be provided in the last subsection, 
and as a step towards the goal, 
we will first describe the Stokes multipliers of $(L_{\rm I})$
which are preserved under the variation of $t$.

\subsubsection{Computing Stokes multipliers of $(L_{\rm I})$ based on Conjecture \ref{conj:hope}}
\label{subsec:linear-Stokes-data}

In what follows, we assume that the claims (i)--(iii) in Conjecture \ref{conj:hope} are true. 
Under the assumption (ii), we have five canonical solutions ${\Psi}^{(j)}_{\pm}$ 
defined in the Stokes region $D_j$ in Figure \ref{fig:mutation} ($j = 0, \pm1, \pm2$ mod $5$).  
Then, we define the Stokes matrix $S_j$ attached to the $j$-th singular direction 
$\arg x = 2\pi j/5$ by\footnote{
It turns out that there is a typo in \cite[eq.\, (2.16)]{IM23}.
The Stokes multipliers in \cite[eq.\,(2.22)--(2.23)]{IM23} 
are computed with the definition \eqref{eq:stokes-labeling} which is employed in this note.
} 
\begin{equation} \label{eq:stokes-labeling}
({\Psi}^{(j)}_{+}, {\Psi}^{(j)}_{-}) = 
({\Psi}^{(j+1)}_{+}, {\Psi}^{(j+1)}_{-}) \cdot {S}_j.
\end{equation}
We also denote by $s_j$ the Stokes multiplier as the non-trivial off-diagonal entry of $S_j$. 

Now, let us recall the proposal of \cite[\S 5]{I19} 
regarding on the computation of the Stokes multipliers $s_j$ based on Conjecture \ref{conj:hope}. 
Let us consider the analytic continuation of ${\mathcal X}_\pm$, 
the Borel resumed perturbative wave function, 
from a region I to II across a Stokes curve $C$, 
just as when we discussed Voros's connection formula in \S \ref{subsec:voros-formula} 
(see Figure \ref{fig:voros1}). 
We also assume that ${\rm Re} \int^{x}_{v} \sqrt{\phi(x)} > 0$ holds on $C$. 
The assertion (iii) in Conjecture \ref{conj:hope} implies that the analytic continuation of
the ${\mathcal X}_+$ across a Stokes curve 
is described by the Borel sum of the formal series \eqref{eq:chi-pm} 
whose coefficients are analytically continued along a detoured path 
such as in Figure \ref{fig:voros2}. 
Then, the path of integration for the resulting term-wise analytic continuation of \eqref{eq:chi-pm} 
is different from the original one, and the difference form a closed cycle on the spectral curve $\Sigma_{P_{\rm I}}$.  
Therefore, we can use the property \eqref{eq:monodromy-chi} to describe 
the aforementioned term-wise analytic continuation along a detoured path. 
For example, if the detoured path and the original path differ by the $B$-cycle, 
then we have the following:
\begin{equation}
\label{eq:connection-formula-chi-pm}
\begin{cases}
\displaystyle
{\mathcal X}^{\rm I}_+(x,t,\nu,\hbar) = {\mathcal X}^{\rm II}_+(x,t,\nu,\hbar) 
+ {\rm i} \frac{{\mathcal Z}(t,\nu+\hbar,\hbar)}{{\mathcal Z}(t,\nu,\hbar)} \,{\mathcal X}^{\rm II}_-(x,t,\nu+\hbar,\hbar),
\\[+.5em]
\displaystyle
{\mathcal X}^{\rm I}_-(x,t,\nu,\hbar) = {\mathcal X}^{\rm II}_-(x,t,\nu,\hbar). 
\end{cases} 
\end{equation}
(See \cite[Proposition 5.5]{I19} for general expression of the 
term-wise analytic continuation along the detoured path.)
Then, the discrete Fourier transform of the above relation leads  
\begin{equation}
\begin{cases}
\displaystyle
{\Psi}^{\rm I}_+(x,t,\nu,\rho,\hbar) = 
{\Psi}^{\rm II}_+(x,t,\nu,\rho,\hbar)
+ {\rm i} \, {\rm e}^{- 2 \pi {\rm i} \rho/\hbar} \, {\Psi}^{\rm II}_-(x,t,\nu,\rho,\hbar), 
\\[+.5em]
\displaystyle
{\Psi}^{\rm I}_-(x,t,\nu,\rho,\hbar) = 
{\Psi}^{\rm II}_-(x,t,\nu,\rho,\hbar).
\end{cases}
\end{equation} 
Thus, we have obtained the non-trivial connection multiplier ${\rm i} \, {\rm e}^{- 2 \pi {\rm i} \rho/\hbar}$.
Similarly, the connection multiplier can also be determined in cases where 
the difference created by the detoured path forms another closed cycle.
As a result, for each $j$, the Stokes coefficient $s_j$ 
can be obtained by applying the aforementioned connection formula 
to all Stokes curves asymptotically approaching in the direction of $\arg x = 2\pi j/5$. 

Omitting the details, the method in \cite[\S 5]{I19} provides the following results 
for the Stokes multipliers when the Stokes graph is as shown in Figure \ref{fig:mutation}:
\begin{align} 
\text{At $t = t_c - {\rm i} \epsilon$} & : \quad
\begin{cases}
s_{-2}  = {\rm i} X_A \\ 
s_{-1}  = {\rm i} (X_A^{-1} - X^{-1}_{A}X^{-1}_{B} + X^{-1}_{B})  \\
s_0 \hspace{+.6em} = {\rm i} X_B \\
s_1 \hspace{+.6em} = {\rm i} (X_B^{-1} - X_A X_B^{-1}) \\
s_2 \hspace{+.6em} = {\rm i} (X_A^{-1} - X_A^{-1} X_B),
\end{cases}
\label{eq:Stokes-B-1}
\\[+1.em]
\text{At $t = t_c + {\rm i} \epsilon$} & : \quad
\begin{cases}
{s}_{-2}  = {\rm i} (X_A - X_A X_B) \\ 
{s}_{-1}  = {\rm i} (X_B^{-1} - X_A^{-1} X_B^{-1}) \\
{s}_0 \hspace{+.6em} = {\rm i} X_B \\
{s}_1 \hspace{+.6em} = {\rm i} (X_A - X_A X_B^{-1} + X_B^{-1}) \\
{s}_2 \hspace{+.6em} = {\rm i} X_A^{-1}, 
\end{cases}
\label{eq:Stokes-B-2}
\end{align}
where $X_A$ and $X_B$ are the exponential Voros periods 
of our non-perturbative wave function: 
\begin{equation}
X_A = {\rm e}^{2 \pi {\rm i} \nu/\hbar}, \quad
X_B = {\rm e}^{2 \pi {\rm i} \rho/\hbar}.
\end{equation}


Thus, the fact described in Part I, 
namely the description of the monodromy (or Stokes) data in terms of the Voros period, 
has been extended to the non-perturbative wave function. 
It is noteworthy that the discrete Fourier transform and the Voros connection formula 
worked in harmony, allowing the Stokes multipliers to be explicitly calculated.
Although the derivation is based on the conjectural arguments,
the resulting Stokes multipliers satisfy 
the following two desired properties that strongly support the validity of our method.

\begin{itemize}
\item 
The consistency conditions
\begin{equation}\label{eq:cyclic}
1 + s_j s_{j-1} + i s_{j+2} = 0 \qquad (j ~{\rm mod}~ 5),
\end{equation}
which guarantees the single-valuedness of the solutions of $(L_{\rm I})$,
are precisely satisfied for both of \eqref{eq:Stokes-B-1} and \eqref{eq:Stokes-B-2}. 

\item
Our method could precisely recover the Kitaev's formula 
on the elliptic asymptotic obtained in \cite{Kitaev}. 
Namely, we can describe the elliptic asymptotic behavior \eqref{eq:boutroux-asymptotic} 
in terms of the Stokes data of $(L_{\rm I})$ by establishing a relation between 
$(\nu, \rho)$ and the Stokes data, thus recovering Kitaev's formula (see see \cite[Remark 5.6]{I19}).

\end{itemize}




\subsubsection{From isomonodromy to resurgent structure}

Now, let us use the results of the previous subsection to derive 
a Stokes jump formula on the perturbative/non-perturbative partition function. 
As is mentioned in \S \ref{subsec:strategy}, we are interested in how the saddle connection on the $B$-cycle 
(or the expected Borel singularity $\oint_B \, y \,{\rm d}x$) contributes to the Stokes jump. 

The main claim of this section is the following.

\begin{conj}[{\cite[\S 2.3]{IM23}}]  ~ 
Let $(\nu^\pm, \rho^\pm)$ 
be two sets of parameters and 
${\mathscr T}^{\pm}(t, \nu^\pm, \rho^\pm, \hbar)$ be the analytic $\tau$-function 
of $(P_{\rm I})$ defined on a domain that contains the point $t = t_c \pm {i} \epsilon$, respectively. 
When $t$ varies and cross the negative real axis, then these analytic $\tau$-functions 
satisfies the following (non-linear) connection formula:  
\begin{equation} \label{eq:Stokes-relation-tau}
{\mathscr T}^{-}(t,\nu^-,\rho^-, \hbar) = 
{\rm e}^{\frac{1}{2 \pi {\rm i}} {\rm Li}_2({\rm e}^{2 \pi {\rm i} \rho^+/\hbar})}
{\mathscr T}^{+}(t, \nu^+, \rho^+, \hbar) 
\end{equation}
with 
\begin{equation} \label{eq:monodromy-symplectomorphism}
( \nu^+, \rho^+ ) = 
\left( \nu^- - \frac{\hbar}{2 \pi {\rm i}} \log(1 - {\rm e}^{2\pi {\rm i} \rho^-/\hbar}),\rho^- \right).
\end{equation}
\end{conj}

Let us explain the derivation of the above-mentioned conjectural formula. 
Since $t$ is the isomonodromic time, the Stokes multipliers corresponding to 
${\mathscr T}^{\pm}(t, \nu^\pm, \rho^\pm, \hbar)$ must be identical if they are related 
by the analytic continuation with respect to $t$.
On the other hand, the Stokes data \eqref{eq:Stokes-B-2}
are obtained from \eqref{eq:Stokes-B-1} by the transformation 
\begin{equation} \label{eq:DDP-B}
(X_A, X_B) \mapsto 
(X_A (1 - X_B), X_B).
\end{equation} 
In fact, this is an example of the {\em Delabaere--Dillinger--Pham formula}
(also known as {\em cluster transformation}, or {\em Kontsevich--Soibelman transformation})
which we mentioned in \S \ref{subsec:comments-wkb}. The prefactor 
$\exp(\frac{1}{2 \pi {\rm i}} {\rm Li}_2({\rm e}^{2 \pi {\rm i} \rho^+/\hbar}))$
on the right hand side of \eqref{eq:Stokes-relation-tau} is related to 
the generating function of the monodromy symplectomorphism \eqref{eq:DDP-B}; 
see \cite{BK19, CLT20} for similar results.  
Based on the above considerations, we arrived at the aforementioned connection formula for the $\tau$-function.

We may simply write the formula by means of the Stokes automorphism as
\begin{equation}
\label{eq:Stokes-auto-tau}
{\mathfrak S} \tau(t,\nu,\rho, \hbar) = 
{\rm e}^{\frac{1}{2 \pi {\rm i}} {\rm Li}_2({\rm e}^{2 \pi {\rm i} \rho/\hbar})}
\tau\left( t, \nu - \frac{\hbar}{2 \pi {\rm i}} \log(1 - {\rm e}^{2\pi {\rm i} \rho/\hbar}) , \rho, \hbar \right). 
\end{equation}
Furthermore, recalling that the $\tau$-function was obtained as the discrete Fourier transform 
of the perturbative partition function, 
we can also derive the connection formula for the perturbative partition function 
by looking at the $0$-Fourier mode of the previous formula 
\eqref{eq:Stokes-relation-tau}--\eqref{eq:monodromy-symplectomorphism}. 
Thus we have 
\begin{conj}[{\cite[\S 2.3]{IM23}; see also \cite{Marino-les-houches-24}}] 
\label{conj:Stokes-partition-function}
The Stokes automorphism associated with the $B$-periods acts on 
the perturbative partition function as follows:
\begin{equation} \label{eq:stokes-auto-pert}
{\mathfrak S} 
Z(t,\nu, \hbar) = 
\exp\left( 
\frac{1}{2 \pi {\rm i}} \, {\rm Li}_2({\rm e}^{- \hbar \partial_\nu}) 
- \frac{ \hbar \partial_\nu}{2 \pi {\rm i}} \log(1 - {\rm e}^{- \hbar \partial_\nu}) \right) 
Z(t,\nu, \hbar).
\end{equation}
\end{conj}

The formula \eqref{eq:stokes-auto-pert} describes all instanton corrections 
to the perturbative partition function explicitly in all order. Namely we can write 
\begin{equation} \label{eq:Stokes-auto-ZTR}
{\mathfrak S} Z(t,\nu, \hbar) = \sum_{n=0}^{\infty} Z^{(n)}(t,\nu, \hbar),
\end{equation}
where the first few terms are explicitly given by
\begin{equation}
Z^{(0)}(t,\nu, \hbar ) = Z(t,\nu, \hbar),  \quad
Z^{(1)}(t,\nu, \hbar) = 
\left( 1 + \frac{\hbar}{2\pi {{\rm i}}} \frac{\partial F}{\partial \nu}(t,\nu - \hbar, \hbar)  \right) 
Z(t,\nu - \hbar , \hbar), \quad \dots.
\label{eq:1-instanton-TR}
\end{equation}
Note also that the Seiberg--Witten relation 
$\partial_\nu F_0 = \oint_{B} y \, {\rm d}x$ makes \eqref{eq:Stokes-auto-ZTR} a well-defined trans-series.  
We can verify that the result \eqref{eq:stokes-auto-pert} precisely agree with 
the multi-instanton results for the topological string obtained in \cite{gm22, gkkm23}
(based on the non-perturbative analysis of {\em holomorphic anomaly equations}\footnote{
An interesting preprint \cite{BGMT24}, that appeared during the writing of this lecture note, 
also discusses the relationship between the $\tau$-function and the holomorphic anomaly equation.
} 
initiated by \cite{CSESV14, CSESV13}), 
where the Stokes constant can be identified with the {\em BPS invariant}, 
as expected\footnote{
Since the BPS invariant for the $B$-cycle is equal to $1$,
it is invisible in the formula \eqref{eq:stokes-auto-pert}. 
See \cite[eq.\,(1.4)]{IM} for example.
}.
We may also observe that the first few terms of the 1-instanton part \eqref{eq:1-instanton-TR}
are consistent with a known connection formula for $(P_{\rm I})$ (see \cite{Kap04} for example). 
These observations strongly support our heuristic derivation of \eqref{eq:Stokes-auto-tau}.

In summary, the connection formulas 
\eqref{eq:Stokes-relation-tau}--\eqref{eq:monodromy-symplectomorphism} 
and \eqref{eq:stokes-auto-pert} are a direct consequence of 
the isomonodoromy property (i.e., {integrability}) 
of the Painlev\'e equation and the DDP formula in exact WKB analysis.

\smallskip
\subsection{Topics related to Part II} \label{subsec:comments-part2}

\begin{itemize}
\item 
Aoki--Kawai--Takei also constructed another type of 2-parameter formal solutions of 
the Painlev\'e equations in \cite{AKT96, KT98-Painleve}, 
where they used the so-called {multiple-scale analysis}\footnote{
See also \cite{Umeta19} for generalization to a class of Painlev\'e hierarchies. 
}.
For example, their 2-parameter solution of $(P_{\rm I})$ takes the form 
\begin{equation} \label{eq:akt-2-parameter-sol}
q(t,\alpha,\beta, \hbar) = 
q_0(t) + \sum_{\ell \ge 1} \hbar^{\ell/2} q_{\ell/2}(t, \alpha, \beta, \hbar),
\end{equation}
where $q_0(t) = \sqrt{-t/6}$, and the terms 
$q_{\ell/2}(t, \alpha, \beta, \hbar)$ are functions of $t$ and $\hbar$ 
containing two free parameters $\alpha, \beta$. 
The first few terms are of the form 
\[
q_{1/2} = \alpha a_{1}(t)\,{\rm e}^{A(t)/\hbar} + \beta a_{-1}(t)\,{\rm e}^{- A(t)/\hbar}, 
\]
\[
q_{1} = \alpha^2 a_{2}(t)\,{\rm e}^{2A(t)/\hbar} + \alpha \beta a_0(t) + \beta^2 a_{-2}(t)\,{\rm e}^{-2A(t)/\hbar}, 
~\dots
\]
with a certain functions $A(t)$, $a_i(t)$ (see \cite[Table 4.6]{KT98} for their explicit expressions). 
These formal solutions were also considered in \cite{GIKM}, 
and studied more extensively in \cite{ASV11,  BSSV22, Del16, vSV22} etc. from the viewpoint of resurgence. 

In \cite{Takei95, Takei98}, Takei derived a formula for Stokes multipliers of $(L_{\rm I})$
by the parameters $(\alpha, \beta)$ under the assumption of a certain summability 
of the 2-parameter solution \eqref{eq:akt-2-parameter-sol}.
On the other hand, as we have seen in \S \ref{subsec:linear-Stokes-data}, 
we also could express the Stokes multipliers 
in terms of the parameters $(\nu, \rho)$ under Conjecture \ref{conj:hope}.
Then, we can propose a conjectural relation between the two sets of free parameters $(\alpha, \beta)$ and $(\nu, \rho)$
by comparing the Stokes multipliers. 
The resulting conjectural relation 
(which is expected to valid on a certain sectorial domain on $t$-plane) is the following (c.f., \cite{BSSV22}):
\begin{equation} \label{eq:2-parmeter-relation}
\begin{cases}
\displaystyle 
\alpha = \frac{\sqrt{\pi}}{\Gamma(\frac{\nu}{\hbar})} \, 2^{\frac{\nu}{\hbar}} \, 
{\rm e}^{-2 \pi {\rm i} (\nu+\rho)/\hbar} \\[+1.5em]
\displaystyle 
\beta = \frac{\Gamma(1 + \frac{\nu}{\hbar})}{\sqrt{\pi}} \,\, 
2^{- 1 - \frac{\nu}{\hbar}} \, {\rm e}^{2 \pi {\rm i} (\nu+\rho)/\hbar}
\end{cases}
\end{equation}
In particular, we may verify that $\nu = 2 \alpha \beta \hbar$ holds; 
this explains why the classical limit of $(L_{\rm I})$ for the Aoki--Kawai--Takei's 
2-parameter solution gives a degenerate elliptic curve (c.f., \cite[\S 4]{KT98}). 
It would be interesting to prove how this relation connects 
the two formal solutions \eqref{eq:akt-2-parameter-sol} 
and \eqref{eq:sol-PI} analytically.




\item 
In \S \ref{subsec:resurgence-painleve}, 
we made numerous considerations that have not been mathematically rigorously proven. 
For example, proving the Borel summability of the perturbative partition function, 
analyzing its Borel singularities, and the convergence of the non-perturbative partition function 
(after taking the Borel), would constitute a significant breakthrough. 
It worth mentioning that the convergence of the Borel transform 
of partition function was proved very recently in \cite{BEG24}. 
From a mathematical viewpoint, in order to state formula \eqref{eq:stokes-auto-pert}, 
it is further necessary to prove that the Borel singularity is mild enough 
to allow the definition of the alien derivative in the first place. 

Providing a rigorous proof is a challenging task, but the connection 
with the Painlev\'e equations might offer some clues for the proof. 
Along the direction, we note that the Borel summability of 
the 0-parameter solution $q(t,\hbar) = \sum_{m \ge 0} \hbar^{m} q_m(t)$ 
(i.e., the one obtained from \eqref{eq:akt-2-parameter-sol} by setting $\alpha = \beta = 0$) 
was rigorously established by Kamimoto--Koike in \cite{KK13}.
More precisely, they prove that the 0-parameter solution is Borel summable if $t$ does not lie on 
``Stokes curve of $(P_{\rm I})$" in the sense of \cite[Definition 4.5]{KT98}.
This fact is valid for all 2nd order Painlev\'e equations $(P_{\rm I})$--$(P_{\rm VI})$.

\item 
In Conjecture \ref{conj:Stokes-partition-function}, we discussed the expected resurgent structure 
for the partition function of the TR. 
Analyzing such resurgent structures in topological string theory, 
not limited to TR, is an extremely important problem, 
as it contributes to elucidating the non-perturbative effects in the models.
Research has accelerated over the past decade, yielding various results 
such as \cite{Alim24, AHT22, ASV11, BT24, CSESV14, CSESV13, gkkm23, gm22} etc.
See also Mari\~no's Les Houches Lecture Note \cite{Marino-les-houches-24}.

\item 
The similarity between the construction of $\tau$-functions 
in Theorem \ref{thm:main-theorem} and Kyiv formula \cite{GIL}
suggests a close relationship among the perturbative partition function 
of TR and the {conformal blocks}. 
Through the Alday--Gaiotto--Tachikawa correspondence (\cite{AGT})
and the {geometric engineering} (\cite{KKV1997}), 
we can understand the relation as a consequence of 
the {remodeling conjecture} of \cite{BKMP, Marino08}, 
which was proved in \cite{EO12, FLZ16}. 
See also \cite{AFKMY, DGH10, KPW10,  MT10}. 
In this direction, recent progress has been made in \cite{AT10, BBCCN, BCU24, CDO24}, 
such as the construction of {Whittaker vectors} based on the {Airy structure} 
(an algebraic formulation of TR proposed by Kontsevich--Soibelman \cite{KS-Airy}). 

However, 
the relation is not fully understood, at least at a mathematically rigorous level, 
for {irregular conformal blocks} discussed in \cite{Nagoya1, Nagoya2, PP23} etc. 
In this case, it is expected that the so-called Argyres--Douglous theory 
corresponds to the gauge theory-side, and there is no proof of the AGT correspondence and 
the remodeling conjecture in this setting. 
However, \cite{BLMST} and recent works \cite{FGFMS, PP23, IILZ25} suggest that 
the previously known dualities can be extended in this class. 
Also, exploring the relationship between the {\em refined topological recursion}, 
as reformulated in recent studies \cite{KO23, Osuga24-1, Osuga24-2}, 
and the conformal blocks with generic central charge should be an intriguing research topic.


Additionally, the Borel sum of perturbative wave function $\chi_\pm$ should be 
related to the irregular conformal block with a degenerate field insertion, 
and the aforementioned connection formula \eqref{eq:connection-formula-chi-pm}
should be compared with those of the conformal blocks. 
In a recent paper \cite{HN24}, the nonabelianization map of $c=1$ Virasoro conformal block
was introduced. Their nonabelianization map is defined by using the spectral network,
and hence, it would be interesting to compare their result and 
our connection formula \eqref{eq:connection-formula-chi-pm}.



\end{itemize}

\bigskip
\ack
The author would like to express gratitude to 
B. Eynard, 
E. Garcia-Failde, 
A. Giacchetto, 
P. Gregori, 
D. Lewa\'nski, 
D. Mitsios 
and 
S. Oukassi
who organize the summer school on 
``Quantum Geometry -- Mathematical Methods for Gravity, Gauge Theories and Non-Perturbative Physics''
held at Les Houches School of Physics. 
He would also like to thank 
T. Aoki, 
I. Coman, 
F. Del Monte, 
H. Fuji, 
K. Hiroe,
S. Hirose, 
N. Iorgov,
H. Iritani, 
S. Kamimoto, 
O. Kidwai, 
T. Koike, 
O. Lisovyy,
M. Mari\~no, 
O. Marchal, 
M. Mulase, 
A. Neitzke, 
Y. Ohyama,
H. Ohta,  
K. Osuga, 
A. Saenz,
H. Sakai,
M.-H. Saito, 
S. Sasaki, 
R. Sch\"afke, 
A. Takahashi, 
T. Takahashi,
Y. Takei, 
Y-M. Takei, 
M. Tanda,
Y. Umeta,
A. Voros,
D. Yamakawa
and 
Y. Zhuravlev 
for their continued support, encouragement, and helpful discussions.
Finally, I would like to acknowledge support from the
Japan Society for the Promotion of Science 
(Grant Number: 24K00525, 23K17654, 22H00094, 21H04994).


\end{document}